\documentclass[conference]{IEEEtran}

\usepackage{times}
\usepackage[numbers]{natbib}
\usepackage{multicol}
\usepackage[bookmarks=true]{hyperref}

\usepackage{graphics} % for pdf, bitmapped graphics files
\usepackage{epsfig} % for postscript graphics files
\usepackage{amssymb}  % assumes amsmath package installed
\usepackage{siunitx}
\DeclareSIUnit\fps{fps}
\usepackage{amsmath,amsfonts}
\usepackage{array}
\usepackage{textcomp}
\usepackage{stfloats}
\usepackage{url}
\usepackage{verbatim}
\usepackage{graphicx}
\usepackage{booktabs}
\usepackage{bm}

\usepackage{threeparttable} % to use table notes
\usepackage{multirow}
\usepackage{color}
\usepackage{float}
\usepackage{balance}

\usepackage{algorithm}
\usepackage{algpseudocode}

\begin{document}

% paper title
\title{
GauSS-MI: Gaussian Splatting Shannon Mutual Information for Active 3D Reconstruction
}

% You will get a Paper-ID when submitting a pdf file to the conference system
% \author{{\color{red}Paper-ID [480]}}

\author{\authorblockN{
Yuhan Xie\authorrefmark{1}, 
Yixi Cai\authorrefmark{2}, 
Yinqiang Zhang\authorrefmark{1}, 
Lei Yang\authorrefmark{3}\authorrefmark{4}, 
and Jia Pan\authorrefmark{1}\authorrefmark{4}
}
\authorblockA{\authorrefmark{1}
School of Computing and Data Science, The University of Hong Kong, Hong Kong SAR, China \\
Email: \{yuhanxie, zyq507\}@connect.hku.hk, jpan@cs.hku.hk
}
\authorblockA{\authorrefmark{2} {Division of Robotics, Perception, and Learning, KTH Royal Institute of Technology, Stockholm, Sweden}\\
Email: yixica@kth.se}
\authorblockA{\authorrefmark{3}
Faculty of Engineering, The University of Hong Kong, Hong Kong SAR, China \\
Email: lyang125@hku.hk}
\authorblockA{\authorrefmark{4}
Centre for Transformative Garment Production, Hong Kong SAR, China }
}

\maketitle

%-----------------------------------% abstract %-----------------------------------%
\begin{abstract}

This research tackles the challenge of real-time active view selection and uncertainty quantification on visual quality for active 3D reconstruction. Visual quality is a critical aspect of 3D reconstruction. Recent advancements such as Neural Radiance Fields (NeRF) and 3D Gaussian Splatting (3DGS) have notably enhanced the image rendering quality of reconstruction models. Nonetheless, the efficient and effective acquisition of input images for reconstruction—specifically, the selection of the most informative viewpoint—remains an open challenge, which is crucial for active reconstruction. Existing studies have primarily focused on evaluating geometric completeness and exploring unobserved or unknown regions, without direct evaluation of the visual uncertainty within the reconstruction model. To address this gap, this paper introduces a probabilistic model that quantifies visual uncertainty for each Gaussian. Leveraging Shannon Mutual Information, we formulate a criterion, Gaussian Splatting Shannon Mutual Information (GauSS-MI), for real-time assessment of visual mutual information from novel viewpoints, facilitating the selection of next best view. GauSS-MI is implemented within an active reconstruction system integrated with a view and motion planner. Extensive experiments across various simulated and real-world scenes showcase the superior visual quality and reconstruction efficiency performance of the proposed system.

\end{abstract}

\IEEEpeerreviewmaketitle

%-----------------------------------% Section 1 %-----------------------------------%
\section{Introduction}

3D reconstruction is attracting increasing interest across various fields, including computer vision\cite{mildenhall2021nerf, kerbl20233d}, manipulation\cite{yang2024one}, robotics\cite{maboudi2023review, zhou2021fuel}, construction\cite{zhang2024global}, etc. 
Recent advancements, such as Neural Radiance Field (NeRF)\cite{mildenhall2021nerf} and 3D Gaussian Splatting (3DGS)\cite{kerbl20233d}, have notably enhanced the visual quality of 3D reconstruction models. 
However, these techniques necessitate the prior acquisition of a significant number of images, which can be laborious, and the extensive sampling of viewpoints may result in redundancy. % labor-intensive
Consequently, a challenging issue arises in effectively and efficiently selecting the viewpoints for image capture, which is also a critical problem for active 3D reconstruction.

To enhance the autonomy of robots and enable them to perform 3D reconstruction tasks in complex environments, there has been a growing focus on active 3D reconstruction in recent years \cite{zhou2021fuel, jin2024gs, tabib2021autonomous}. 
In the active 3D reconstruction process, at each decision step, the agent must utilize a series of past observations to actively determine the next viewpoint for capturing new observation, thus gradually accomplishing the reconstruction task. 
The efficient selection of viewpoints is particularly crucial in this process due to limited onboard resources such as battery power, memory, and computation capability. 
Previous studies on active 3D reconstruction have primarily relied on evaluating volumetric completeness to explore all unknown voxels in the environment \cite{isler2016information, zhou2021fuel, tabib2021autonomous} or assessing surface coverage completeness \cite{cao2020hierarchical, feng2024fc}. 
% These approaches do not directly consider visual quality. 
These approaches overlook the visual quality.
By utilizing these indirect metrics, the resulting visual fidelity of the reconstruction model cannot be guaranteed.
Advanced by radiance field rendering methods\cite{kerbl20233d, mildenhall2021nerf}, recent works have attempted to quantify visual uncertainty to directly evaluate the visual quality of reconstruction models \cite{shen2021stochastic, goli2024bayes, jiang2025fisherrf}.

Despite these efforts, effectively and efficiently assessing and optimizing visual quality in active 3D reconstruction remains challenging. 
To address this, three core issues must be resolved. 
Firstly, a robust mathematical model is necessary to quantify the information obtained from each measurement, specifically the observed image. 
This model can serve as a reconstruction completeness metric for visual fidelity. 
Secondly, a metric is needed to measure the expected information from novel viewpoints without a prior, which can facilitate the selection of the next viewpoint in the active reconstruction process. 
Lastly, a comprehensive active reconstruction system is required to autonomously identify a reasonable next viewpoint with the highest expected information.
The system should then enable the agent to navigate to the selected viewpoint, capture new data, and iteratively advance the reconstruction process.

To overcome the aforementioned challenges, this paper proposes a novel view selection metric based on a visual uncertainty quantification method, from which we develop a novel active 3D reconstruction system. 
We first introduce a probabilistic model that integrates the measurement model with image loss to quantify the observed information for each spherical Gaussian in 3D Gaussian Splatting. 
Based on Shannon Mutual Information theory, we leverage the probabilistic model to establish the mutual information between the reconstruction model and observation viewpoint, which measures the expected information gained from an arbitrary viewpoint for the current reconstruction model.
This mutual information function is termed Gaussian Splatting Shannon Mutual Information (GauSS-MI), enabling real-time visual quality assessment from novel viewpoints without a prior. 
The GauSS-MI is implemented and integrated into a novel active 3D Gaussian splatting reconstruction system featuring a view and motion planner that determines the next best view and optimal motion primitive. 
Extensive experiments, including benchmark comparisons against state-of-the-art methods, validate the superior performance of the proposed system in terms of visual fidelity and reconstruction efficiency. 
The implementation of the proposed system is open-sourced on Github\footnote{\url{https://github.com/JohannaXie/GauSS-MI} } to support and advance future research within the community.
%to benefit the community for future research.

The main contributions of our work are summarized below:
\begin{itemize}
{   %% this be concise, details in the former paragraph
    \item A probabilistic model for the 3D Gaussian Splatting map to quantify the image rendering uncertainty. %observed information.
    \item A novel Gaussian Splatting Shannon Mutual Information (GauSS-MI) metric for real-time assessment of visual mutual information from novel viewpoints. 
    \item An active 3D Gaussian splatting reconstruction system implementation based on GauSS-MI. % with a view and motion planner.
    \item Extensive benchmark experiments against state-of-the-art methods demonstrate the superior performance of the proposed system in terms of visual fidelity and reconstruction efficiency.  
}
\end{itemize}

\begin{figure*}[t] 
    \centering
    \includegraphics[width=\linewidth]{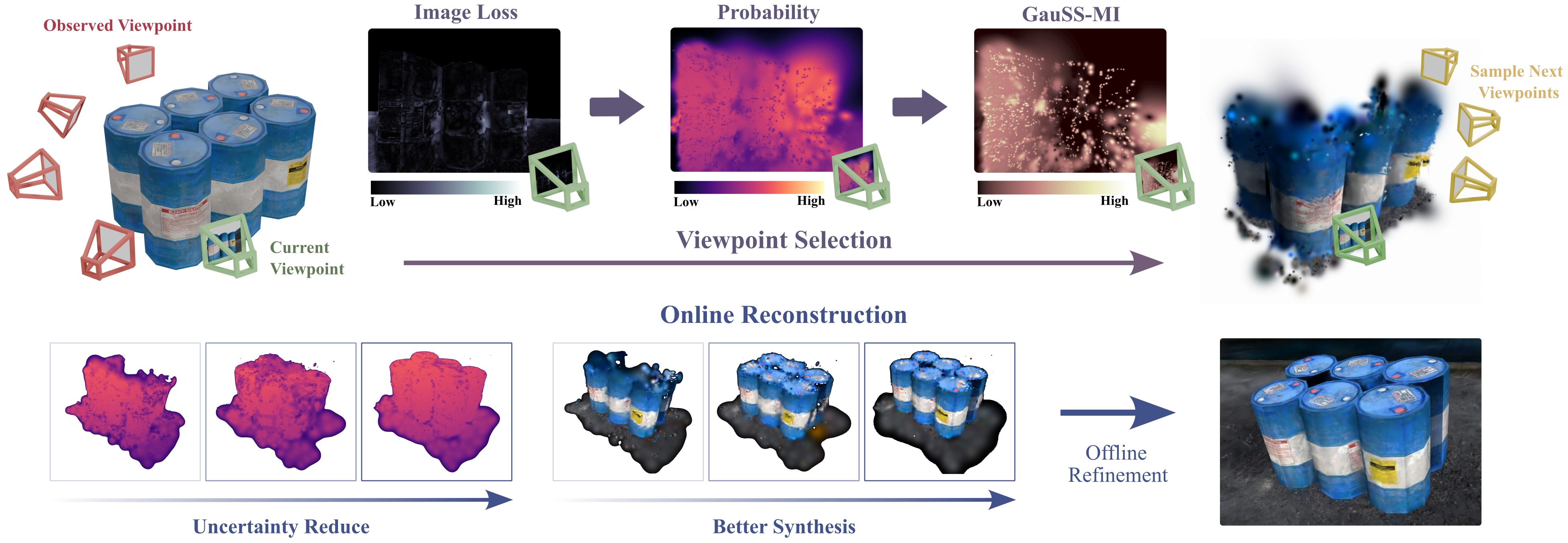}  
\caption{Illustration of the proposed Gaussian Splatting Shannon Mutual Information (GauSS-MI) method. 
\textbf{Upper part}: 
At each active reconstruction step, once a new observation is obtained, the 3D Gaussian Splatting (3DGS) map is updated and optimized by minimizing the image loss between observed images and the map.
To quantify visual uncertainty, we construct a probabilistic model for each 3D Gaussian ellipsoid by mapping residual image loss onto the 3DGS map.
Using this model, we define GauSS-MI, a metric that estimates mutual information between the reconstruction model and a viewpoint. 
GauSS-MI enables real-time visual quality assessment from novel viewpoints without a prior, facilitating the selection of the next-best-view to effectively reduce map uncertainty. 
% The next best view is then selected by GauSS-MI to reduce the uncertainty of the current map effectively. 
\textbf{Lower part}: 
The active reconstruction process iterates and decreases visual uncertainty, resulting in a high visual fidelity 3D reconstruction result.
% The process iterates and the visual uncertainty of the reconstruction model decreases, resulting in a high visual fidelity 3D reconstruction model.
}
    \label{fig_main}
\end{figure*}

%-----------------------------------% Section 2 %-----------------------------------%

\section{Related Work}

The evolution of mapping representations in 3D reconstruction has driven significant advancements in active reconstruction methodologies. A key distinction between active and passive reconstruction lies in the process of active view selection. 
In this section, we first review active view selection strategies across various mapping representations. We then present a detailed review of uncertainty quantification techniques employed in information-based approaches.

% \end{itemize}
\subsection{Active View Selection for 3D reconstruction}
The first branch of research in active 3D reconstruction focuses on geometric reconstruction, utilizing occupancy-based representations \cite{hornung2013octomap}. In this domain, a commonly employed strategy for determining the next best viewpoint involves constructing and evaluating frontiers, which indicate the boundary between mapped and unmapped areas\cite{yamauchi1997frontier, cieslewski2017rapid, zhou2021fuel}. 
Additionally, researchers have incorporated information-theoretic approaches, considering metrics such as information gain \cite{isler2016information, lu2024semantics} and mutual information \cite{zhang2020fsmi, julian2014mutual, tabib2021autonomous}, to maximize the information observed at subsequent viewpoints. 
Recent studies have also explored surface coverage in active reconstruction, with or without prior knowledge of the environment\cite{feng2023predrecon,feng2024fc,zhang2024falcon}, to enhance reconstruction efficiency. However, methods within this branch mainly rely on occupancy-based map information, which does not inherently ensure high visual quality in the resulting reconstructions.

Recent advancements in radiance field representations, such as Neural Radiance Fields (NeRF) \cite{mildenhall2021nerf} and 3D Gaussian Splatting (3DGS) \cite{kerbl20233d}, have significantly enhanced visual quality in 3D reconstruction, sparking interest in their application to active reconstruction for even higher visual fidelity. 
However, early research primarily rely on geometric information derived from occupancy maps for viewpoint selection, rather than directly leveraging the rich visual information inherent in radiance field maps. 
For instance, \citet{yan2023active} explored geometric completeness in NeRF by evaluating information gain based on volumetric data. \citet{li2024activesplat} introduced ActiveSplat, a method that optimizes environment coverage to achieve high visual quality using 3DGS. In NARUTO \cite{feng2024naruto}, Feng et al. considered implicit uncertainty in geometric information for active reconstruction with 3DGS. 

Considering the implicit model in NeRF, researchers have explored neural network-based approaches for evaluating implicit uncertainty in visual quality\cite{lee2023so, pan2022activenerf, ran2023neurar, shen2021stochastic}. However, while these methods successfully evaluate uncertainty in NeRF for next viewpoint selection, their effectiveness depends heavily on the availability of high-quality datasets for training the evaluation neural networks.

In contrast to NeRF’s implicit modeling approach, 3DGS provides an explicit representation of the environment using a collection of spatially distributed spherical Gaussians. 
This explicit representation facilitates the direct evaluation of uncertainty in visual quality. In GS-Planner \cite{jin2024gs}, image loss is directly incorporated into the occupancy map, enabling the evaluation of both geometric and photometric uncertainty. 
\citet{jiang2025fisherrf} introduced FisherRF, which leverages the Fisher information matrix to quantify the parameter uncertainty in radiance maps. 
Building on this, \citet{xu2024hgs} extended the GS-Planner framework by integrating FisherRF and geometric completeness for more comprehensive uncertainty evaluation. 
ActiveGS by \citet{jin2025activegs} proposed a model for evaluating the confidence of each Gaussian, which is subsequently used for viewpoint selection in active reconstruction. 
In a recent study, \citet{chen2025activegamer} proposed ActiveGAMER with a silhouette-based information gain to enhance both geometric and photometric reconstruction accuracy.
Our method also employs 3DGS as the scene representation for active reconstruction. To enhance the evaluation of observed information in the Gaussian map, we propose a more comprehensive probabilistic model that accounts for both reconstruction loss and sensor measurement noise.

\subsection{Information-theoretic View Uncertainty Quantification}

Information theory offers a robust mathematical framework for active 3D reconstruction, enabling the selection of viewpoints that maximize expected information and consequently reduce the map entropy. 
% As the occupancy grid maps are typically represented in probability maps\cite{julian2014mutual}, which naturally include the observed information and uncertainties. 
Occupancy grid maps, typically represented as probability models\cite{julian2014mutual}, inherently capture observed information and associated uncertainties. 
Consequently, information-theoretic approaches can be directly integrated into occupancy-based active reconstruction.
% Information-theoretic approaches have been extensively investigated in occupancy-based active reconstruction. 
One common method constructs information gain from each observation by occupancy probabilities \cite{isler2016information, lu2024semantics}. 
Alternatively, mutual information has been extensively studied for its ability to quantify the mutual information between a map and observations. 
Shannon Mutual Information (SMI) has been proven to provide guarantees for comprehensive exploration of the environment \cite{julian2014mutual}, demonstrating its theoretical effectiveness and completeness in active reconstruction tasks. 
However, its practical application is hindered by substantial computational overhead when applied to occupancy mapping, as the computational complexity scales quadratically with the spatial resolution of the map and linearly with the numerical integration of range measurements.
To overcome these limitations, \citet{charrow2015information} introduced the Cauchy-Schwarz Quadratic Mutual Information (CSQMI) metric, which enables analytical computation of measurement integration and reduces computational complexity to linear scaling with the map’s spatial resolution. 
Subsequent studies have demonstrated the efficiency of CSQMI in real-time robotic systems for active reconstruction tasks \cite{charrow2015theoretic, nelson2015information, tabib2016computationally}. 
To preserve the theoretical guarantee of SMI, \citet{zhang2020fsmi} proposed the Fast Shannon Mutual Information (FSMI) algorithm, which significantly enhances the computational efficiency of mutual information evaluation compared to the original SMI algorithm \cite{julian2014mutual} by analytically evaluating integrals.

Information-theoretic uncertainty quantification for radiance field-based approaches can be broadly divided into two categories. 
The first category learns an implicit probability model to estimate information gain for novel viewpoints \cite{shen2021stochastic}. 
The second employs the Fisher information matrix, derived from the rendering loss, to quantify information gain\cite{goli2024bayes, jiang2025fisherrf}. 
Among these, FisherRF, a recent method based on the Fisher information matrix, extends its applicability to 3D Gaussian Splatting (3DGS) \cite{jiang2025fisherrf}. 
However, FisherRF focuses primarily on next-best-view (NBV) selection, neglecting the real-time demands of active reconstruction. 
To overcome these limitations, we propose a probabilistic model for 3DGS based on the rendering quality. 
Utilizing the computationally efficient SMI method, which jointly accounts for uncertainties in the reconstructed map and measurements, we introduce Gaussian Splatting Shannon Mutual Information (GauSS-MI), a novel method for quantifying visual uncertainties.
Additionally, we develop an active reconstruction system based on GauSS-MI, which achieves high visual fidelity with real-time performance requirements.

%-----------------------------------% Section 3 %-----------------------------------%

\section{Overview}
This paper introduces Gaussian Splatting Shannon Mutual Information (GauSS-MI) as a metric for efficient next best view selection in high-visual fidelity active reconstruction. 
The proposed method is illustrated in Figure~\ref{fig_main}. 
% During the active reconstruction pro
At each active reconstruction step, we assume that we have a set of previous observations and a set of next viewpoint candidates.
Our goal is to devise an effective metric for the next best view selection leveraging the available observed information.

During each active reconstruction step, a new observation is obtained, and the 3D Gaussian Splatting (3DGS) map is updated by extending and initializing new Gaussians based on this new observation. 
% first extends and initializes new Gaussians based on the new observation.  
Subsequently, the overall 3DGS map undergoes iterative optimization based on the loss between the observed images and the map.
To quantify the visual uncertainty, we compute the remaining image loss between the current observation and the optimized map.
% Utlize the image loss, we 
By mapping the image loss onto the 3DGS map, we construct a probabilistic model for each 3D Gaussian. 
% We map the image loss onto the 3DGS map, and construct a probabilistic model for each 3D Gaussian.
Subsequently, based on Shannon Mutual Information theory, we leverage the probabilistic model to establish the mutual information between the reconstruction model and a viewpoint. 
This mutual information function is referred to as Gaussian Splatting Shannon Mutual Information (GauSS-MI), enabling real-time visual quality assessment from novel viewpoints without a prior. 
The next best view is then selected using GauSS-MI, to effectively reduce the uncertainty of current map.
The iterative process leads to a decrease in visual uncertainty within the reconstruction model, yielding a high visual fidelity 3D reconstruction result.
% The process iterates and the visual uncertainty of the reconstruction model decreases, resulting in a high visual fidelity 3D reconstruction model.

The derivation of GauSS-MI is elaborated in Section~\ref{sec_method}, and the system implementation details are presented in Section~\ref{sec_sys_imple}.

%-----------------------------------% Section - methodology %-----------------------------------%
\section{Methodology}
\label{sec_method}

This section presents the probabilistic model for 3D Gaussian Splatting (3DGS) in visual uncertainty quantification, followed by the formulation of Gaussian Splatting Shannon Mutual Information (GauSS-MI) for view selection. 
The main notations of this section are listed in Table~\ref{tab_notation_map}.

\begin{table}[t]
\centering
\caption{Main Notations for GauSS-MI}
\label{tab_notation_map}
\begin{tabular}{@{}ll@{}}
    \toprule
Notations & Explanation \\ 
    \midrule
    %% 1 - 3DGS map
    $\mathcal{G}$         & 3D Gaussian splatting map. \\
    $\mathcal{W}$           & The world frame. \\
$\mathcal{N}$           & A series of ordered Gaussians along a camera ray. \\
    $\bm\mu$             & Position of a Gaussian. \\
    $c$             & Color of a Gaussian. \\
    $\alpha$             & Opacity of a Gaussian. \\
    $\bm\sigma$         & Camera pose, or a viewpoint.       \\
    $T$             & Cumulative transmittance of a Gaussian. \\
    ${C}, \hat{C}$             & Rendered color and observed color. \\
    ${D}, \hat{D}$             & Rendered depth and observed depth. \\
    
    %% 3 - info
    $z, Z$				& Random variable and realization of an observation. \\
    $m, M$             &  Random variable and realization of luminance for a pixel. \\
    $P(r)$             & \textit{Real} probability of a Gaussian. \\
    $o, l$             & Odds ratio and log odds of a Gaussian. \\
    $\delta$             & Inverse sensor model. \\
    $L$             & Loss image between the map and the observation. \\
    $\lambda$             & Hyperparameters. \\
    $H$             & Entropy of the map. \\
    $I$             & Mutual Information between the map and the observation. \\
    
    %% 2 - 
    ${(\cdot)}^{[i]}$           & Iteration of Gaussians. \\
    ${(\cdot)}^{[j]}$           & Iteration of the measurement beams or pixels. \\
    ${(\cdot)}_k$           & Property based on observation at time $k$. \\
    ${(\cdot)}_{1:k}$           & Property based on the observations from start to time $k$. \\
    \bottomrule
\end{tabular}
\end{table}

\subsection{3D Gaussian Splatting Mapping}\label{sec_sub_map}

The proposed system reconstructs the scene by 3DGS, utilizing a collection of anisotropic 3D Gaussians, represented by $\mathcal{G}$.
Each 3D Gaussian $i$ contains the properties of mean $\bm \mu^{[i]}_\mathcal{W}$ and covariance $\bm \Sigma^{[i]}_\mathcal{W}$, representing the geometrical position and ellipsoidal shape in the world frame $\mathcal{W}$, and also optical properties including color $c^{[i]}$ and opacity $\alpha^{[i]}$. 
By splatting and blending a series of ordered Gaussians $\mathcal{N}$, the color ${C}^{[j]}$ and depth ${D}^{[j]}$ for each pixel are synthesized as
\begin{equation}\label{eq_color_raster}
{C}^{[j]}
= \sum_{i \in\mathcal{N}}{ c^{[i]}T^{[i]} } 
= \sum_{i \in\mathcal{N}}{c^{[i]}\alpha^{[i]} \prod_{n = 1}^{i-1} (1-\alpha^{[n]})}
\end{equation}
\begin{equation}\label{eq_depth_raster}
{D}^{[j]}
= \sum_{i \in\mathcal{N}}{ d^{[i]}T^{[i]} } 
= \sum_{i \in\mathcal{N}}{d^{[i]}\alpha^{[i]} \prod_{n = 1}^{i-1} (1-\alpha^{[n]}) }
\end{equation}
where $d^{[i]}$ represents the distance from camera pose $\bm\sigma$ to the position $\bm \mu^{[i]}_\mathcal{W}$ of Gaussian $i$ along the camera ray. 
We denote
\begin{equation}
T^{[i]}= \alpha^{[i]} \prod_{n = 1}^{i-1} (1-\alpha^{[n]})
\end{equation}
as the cumulative transmittance of Gaussian $i$ along the ray. 

At each reconstruction step, the 3D Gaussians are extended and initialized using the collected RGB-D image and estimated camera pose \cite{Matsuki2024CVPR}. 
Then the Gaussians iteratively optimize both their geometric and optical parameters to represent the captured scene with high visual fidelity.

\subsection{3D Gaussian probability} \label{subsec_3dgs_prob}
% {\color{red} correct $->$ reliable}

To model the information obtained from the 3DGS map $\mathcal{G}$ by a random observation $z$, we first construct a random variable $r$ for each Gaussian.
As we are going to optimize the rendering result, we define the probability of a 3D Gaussian $i$ is \textit{reliable} for rendering as $P(r^{[i]}) \in(0,1)$. 
Then, the probability of the 3D Gaussian $i$ is \textit{unreliable} for rendering is $P(\bar r^{[i]}) = 1-P(r^{[i]})$. 
Additionally, we denote the odds ratio $o^{[i]}\in(0,+\infty)$ and log odds $l^{[i]}\in(-\infty,+\infty)$ of a Gaussian by
\begin{equation}\label{eq_pol_define}
    l^{[i]} := \log(o^{[i]}) := \log(\frac{ P(r^{[i]}) }{  P(\bar r^{[i]})  })
        = \log(\frac{  P(r^{[i]})  }{  1-P(r^{[i]}) })
\end{equation}

We assume each probability of the 3D Gaussian is independent. 
At the initial stage of the mapping, we assume that the agent has no prior information on the environment, i.e.,
\begin{equation}\label{eq_imple_init_prob} 
    P_0(r^{[i]}) = P_0(\bar r^{[i]}) =0.5
 \quad\forall i\in\mathcal{G}
\end{equation}
Once a new observation $Z_k$ is obtained at time $k$, the standard binary Bayesian filter can be used to update the probability
\begin{equation}
\begin{aligned}    \label{eq_prob_update} 
    o^{[i]} (Z_{1:k}) :&=
    \frac{  P(r^{[i]}|Z_{1:k})  }{  P(\bar r^{[i]}|Z_{1:k})}    
    \\
    & = 
        \frac{P(r^{[i]}|Z_k)}         {P(\bar r^{[i]}|Z_k)} 
        \frac{P(r^{[i]}|Z_{1:k-1})}   {P(\bar r^{[i]}|Z_{1:k-1})}
    \\
    & = \delta^{[i]} (Z_k)    o^{[i]} (Z_{1:k-1})
\end{aligned}
\end{equation}
where $P(r^{[i]}|Z_k)$ is the \textit{reliable} probability of Gaussian $i$ under the observation $Z_k$.
We refer to $P(r^{[i]}|Z_k)$ as the inverse sensor model, thereby $\delta^{[i]} (Z_k)$ is the odds ratio of the inverse sensor model, which will be constructed and used for updating the \textit{reliable} probability $P(r^{[i]}|Z_{1:k})$. 
% In the remainder of this section, we use $o^{[i]}_{1:k}$ as a shorthand of $o^{[i]} (Z_{1:k})$, referring to the odds ratio for Gaussian $i$ based on the observations from start to time $k$. 
We further use $o^{[i]}_{1:k}$ and $l^{[i]}_{1:k}$ as a shorthand of $o^{[i]} (Z_{1:k})$ and $l^{[i]} (Z_{1:k})$ respectively, referring to the odds ratio for Gaussian $i$ based on the observations from start to time $k$.

Given the observation $Z_k$, we construct the $P(r^{[i]}|Z_k)$ as
\begin{equation}    \label{eq_inv_sensor_p}
    P(r^{[i]}|Z_k) = \frac{1}{(\lambda_L L_k)^{\lambda_TT^{[i]}} +1}
\end{equation}
Therefore, the odds ratio of inverse sensor model $\delta^{[i]} (Z_k)$ can be derived as
\begin{equation}    \label{eq_inv_sensor}
    \delta^{[i]} (Z_k) 
        = \frac{  P(r^{[i]}|Z_k)  }{  1- P( r^{[i]}|Z_k)}
        = (\lambda_L L_k)^{-\lambda_TT^{[i]}}
\end{equation}
where $\lambda_L, \lambda_T>0$ are hyperparameters. 
$L_k$ denotes the loss between the observation $Z_k$ and the map, i.e., a loss image between the observed groundtruth image and the rendered image. We compute the loss image by
% $L_k$ is a loss image between the observation $Z_k$ with the 3DGS map, i.e., a loss image between the observed groundtruth image and the rendered image. We compute the loss image by
\begin{equation}    \label{eq_loss_image}
    L_k = \lambda_c\|C-\hat{C}_k\| + (1-\lambda_c)\|D-\hat{D}_k\|
\end{equation}
where $C, D$ denote the rendered color and depth images from the reconstructed 3DGS map, $\hat{C}_k$, $\hat{D}_k$ are the groundtruth color and depth images from observation $Z_k$.

\begin{figure}[t] 
    \centering
    \includegraphics[width=\linewidth]{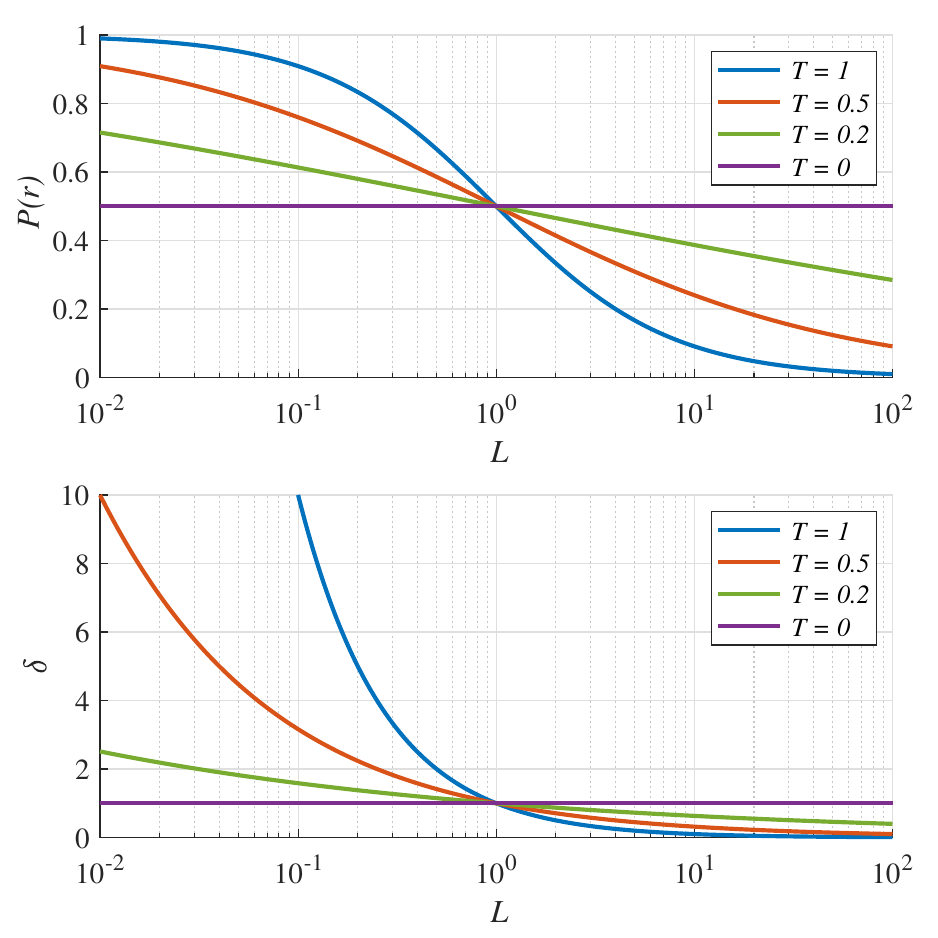}
    \caption{
        Inverse sensor model visualization.
        The hyperparameters $\lambda$ are omitted in the figure for simplicity.
    }
    \label{fig_prob_model}
\end{figure}

As the 3DGS map optimizes the Gaussians by minimizing the image loss, we use this loss to construct the inverse sensor model, and the cumulative transmittance to regulate the update rate. 
We further visualize the inverse sensor model \eqref{eq_inv_sensor_p}\eqref{eq_inv_sensor} in Figure~\ref{fig_prob_model} to illustrate the probability update. 
The Gaussians associated with observation $Z_k$ have $T^{[i]}>0$, resulting in the inverse sensor model $P(r^{[i]}|Z_k)$ and $\delta^{[i]} (Z_k)$ being monotonically decreasing with loss $L_k$. 
This suggests that lower loss $L_k$ corresponds to a higher \textit{reliable} probability of Gaussian $P(r^{[i]}|Z_k)$. 
Additionally, a lower cumulative transmittance $T^{[i]}$ implies less impact of Gaussian $i$ on observation $Z_k$. 
Consequently, a smaller $T^{[i]}$ results in less observed information within the inverse sensor model, e.g., when $T^{[i]}=0$, we have $P(r^{[i]}|Z_k)=0.5$ and $\delta^{[i]}(Z_k)=1$.

To accelerate computation, in implementation, we update the probability $P(r^{[i]}|Z_{1:k})$ by computing log odds $l_{1:k}^{[i]}$.
Take $\log$ of \eqref{eq_prob_update} and substitute \eqref{eq_inv_sensor},
\begin{equation}\label{eq_imple_update_prob} 
    l_{1:k}^{[i]} = 
    -\lambda_TT^{[i]}\log{\lambda_L L_k}    +
    l_{1:k-1}^{[i]}
\end{equation}
Therefore, the log odds of inverse sensor model can be computed by rasterizing the mapping loss $L_k$ as \eqref{eq_color_raster}.
% At each observation step, we compute the mapping loss $L_k$ and rasterize it as \eqref{eq_color_raster}
The probability update algorithm is summarized in Algorithm~\ref{alg_prob_update}.

\begin{algorithm}[t]
\caption{Probability Update}\label{alg_prob_update}

\begin{algorithmic}[1]
\Require 3DGS map $\mathcal{G}$; Observations $\hat{C}_k, \hat{D}_k$ from $\bm\sigma_k$
    \State $\mathcal{N} \gets \mathtt{SortGaussians} (\bm\mu, \bm\sigma_k)$
    \State $C, D \gets\mathtt{ImageRender} (c, \alpha, \bm\mu, \mathcal{N})$
    \State $L_k = \lambda_c||C-\hat{C}_k|| + (1-\lambda_c)||D-\hat{D}_k||$  \Comment{Loss image}
    \State $l_k =0$
    \For {$j \gets  1$ \textnormal{to} $n_z$}              \Comment{Per pixel}
        \State $T^{[j]} = 1$
        \For {$i \gets 1$ \textnormal{to} $\mathcal{N}^{[j]}$}      \Comment{Per gaussian}
            \State $T^{[i]} =  \alpha^{[i]}T^{[j]}$             \Comment{Gaussian's transmittance}
            \State $l_k^{[i]} \mathrel{{-}{=}} \lambda_TT^{[i]}\log{\lambda_L L_k^{[j]}}$ \Comment{Log odds of \eqref{eq_inv_sensor_p}}
            \State $T^{[j]} =  (1-\alpha^{[i]})T^{[j]}$         \Comment{Update pixel transmittance}
        \EndFor
    \EndFor
    \State $l_{1:k} = l_k+ l_{1:k-1} $
    \State $P_{1:k} \gets \mathtt{InvertLogodds} (l_{1:k})$        \Comment{Update probability}
    \State \textbf{return} $P_{1:k}$
\end{algorithmic}
\end{algorithm}

\subsection{Gaussian Splatting Shannon Mutual Information }\label{subsec_GSSMI}
Based on the proposed probability model and Shannon Mutual Information theory, we then construct the Gaussian Splatting Shannon mutual information (GauSS-MI) for visual quality assessment of novel viewpoints.

% \subsubsection{Shannon Mutual Information}
Given the previous observations $Z_{1:k-1}$, we are interested in minimizing the expected uncertainty, i.e., conditional entropy, of the map after receiving the agent's next observation $z_k$. 
In information theory, the conditional entropy relates to Mutual Information (MI) by 
\begin{equation}\label{eq_Ent_MI_full}
    H(r|z_{k}, Z_{1:k-1}) = H(r|Z_{1:k-1}) - I(r; z_{k}|Z_{1:k-1})
\end{equation}
To minimize the conditional entropy $H(r|z_{k}, Z_{1:k-1})$ is to maximize the MI $I(r; z_{k}|Z_{1:k-1})$. 
Note here that we use $z_k$ and $Z_k$ to distinguish random variable and realized variable for the observation at time $k$. 

As we assume that the previous observations $Z_{1:k-1}$ are given and try to compute the MI for the new observation $z_k$,
in the subsequent of this subsection, we omit the probability condition $Z_{1:k-1}$ and simplify $z_k$ into $z$.
Therefore, the \eqref{eq_Ent_MI_full} can be simplified as
\begin{equation}\label{eq_Ent_MI_simp}
    H(r|z) = H(r) - I(r; z)
\end{equation}

As $z$ is a random variable with independence among elements, the total MI can be expressed as the summation of $ I(r ; z^{[j]})$ between $z^{[j]}$ over all measurement beams $j\in\{1,\cdots,n_z \}$ \cite{julian2014mutual}.
\begin{equation}\label{eq_I_decomp}
    I(r; z) = \sum_{j=1}^{n_z} I(r ; z^{[j]})
    =\sum_{j=1}^{n_z} 
    \sum_{i \in\mathcal{N}^{[j]}}{
     I(r^{[i]} ; z^{[j]})  T^{[i]}
    }
\end{equation}
Here, the measurement beams $j\in\{1,\cdots,n_z \}$ refer to each picture pixel. 
% where $\mathcal{N}^{[j]}$ denotes the set of Gaussians that can be seen in the beam $z^{[j]}$.

From information theory \cite{cover1991information, julian2014mutual}, the mutual information between two random variables is defined and can be organized as
\begin{equation}
\begin{aligned}    \label{eq_MI_definition} 
    I(r^{[i]} &; z^{[j]})   
\\ :=&
    P(r^{[i]}, z^{[j]}=Z) 
    \log(\frac  { P(r^{[i]}, z^{[j]}=Z )  }
                { P(r^{[i]} )  
                    P(z^{[j]}=Z  ) } )
\\ = &
    P(z^{[j]}=Z)
% \\  &
    P(r^{[i]}|z^{[j]}=Z  )
    \log(\frac{ P(r^{[i]}|z^{[j]}=Z) }
                { P(r^{[i]} )  } )
\\ = &
    P(z^{[j]}=Z)f(\delta^{[i]}(Z), o^{[i]}_{1:k-1})
\end{aligned}
\end{equation}
where $P(z^{[j]}=Z)$ is only related to the observation, which is referred to as the measurement prior. 
% $f(\delta^{[i]}(z), o_{k-1}^{[i]})$ 
$f(\delta^{[i]}(Z), o^{[i]}_{1:k-1})$ 
can be derived and written in shorthand as
\begin{equation}    \label{eq_info_gain}
    f(\delta, o) = 
    \frac{ o }{o+\delta^{-1}} 
    \log(\frac{o+1}{o+\delta^{-1}}) 
\end{equation}
The detailed derivation is presented in the Appendix~\ref{apn_sec_infogainf}. 
The function $f(\delta, o)$ can be interpreted as an information gain function.

We define the mutual information \eqref{eq_MI_definition} between the 3DGS map and the observation as Gaussian Splatting Shannon Mutual Information, GauSS-MI.

\subsection{Computation of Expected GauSS-MI}
We further derive the computation of the expected mutual information \eqref{eq_MI_definition} for random viewpoints. 

\subsubsection{Measurement prior}
% For the measurement prior probability $P(z)$, 
We refer to the noise model of RGB camera in \cite{foi2008practical}, in which the expectation of the measurement noise is related to luminance.
Thus, we construct the measurement prior $P(z)$ for each pixel $j$ as
% As our measurement sensor is the RGB camera, the measurement beams $j\in\{1,\cdots,n_z \}$ refer to each picture pixel. 
% The measurement prior for each pixel $j$ can be expressed as functions of the sensor model \cite{foi2008practical}.
\begin{equation}    \label{eq_measure_prior}
    P(z^{[j]} ) = \sum_{m^{[j]}=0}^{255} P(z^{[j]}|m^{[j]})P(m^{[j]})
\end{equation}
% \begin{equation}    
%     P(z^{[j]} ) = P(z^{[j]}|j)P(j)
% \end{equation}
where $P(z^{[j]}|m^{[j]})$ is the prior probability distribution of the sensor with respect to luminance $m\in\{0,\cdots,255\}$.
To compute the expected measurement prior, we define $P(m^{[j]})$ as
\begin{equation}
P(m^{[j]}) =
\begin{cases}
    1   & \text{for}\, m^{[j]}=M^{[j]} \\
    0   & \text{otherwise}
\end{cases}
\end{equation}
where $M^{[j]}$ is the pixel's expected luminance, which can be computed from the expected RGB color $(R, G, B)$ by the luminance formula $M = 0.299R + 0.587G + 0.114B$.
Thus the measurement prior \eqref{eq_measure_prior} can be simplified as
\begin{equation}\label{eq_exp_pz}
    P(z^{[j]} ) =  P(z^{[j]}|M^{[j]})
\end{equation}

\subsubsection{Information gain function}

As there are no observations from random viewpoints, computing the loss image $L_k$ for $\delta(Z_k)$ is infeasible; thus, an expectation of $L_k$ is required.
We expect that the rendering result after reconstruction is \textit{reliable}, i.e., there is no loss between groundtruth and the 3DGS map $\mathcal{G}$. 
Thus, we assume $L_k=0$ so that $\delta^{-1}=0$ in $f(\delta,o)$.
Then the information gain function \eqref{eq_info_gain} can be derived as,
\begin{equation}    \label{eq_info_gain_comp}
    f^{[i]}
    = \log(  \frac{o^{[i]}+1}{o^{[i]}}  ) 
    = -\log(  P(r^{[i]}))
\end{equation}
The equation shows that when the \textit{reliable} probability $P(r^{[i]})$ is low, the information gain function $f$ will be high, consistent with the intuition of information gain.

Overall, integrating \eqref{eq_I_decomp}\eqref{eq_MI_definition}\eqref{eq_exp_pz}\eqref{eq_info_gain_comp}, the expected GauSS-MI can be computed as
\begin{equation}
\begin{aligned}    \label{eq_MI_compute} 
    I(r; z) = &
    \sum_{j=1}^{n_z} 
    \sum_{i \in\mathcal{N}^{[j]}}{
     I(r^{[i]} ; z^{[j]})  T^{[i]}
    }
\\ = &
    \sum_{j=1}^{n_z} P(z^{[j]}|M^{[j]})
    \sum_{i \in\mathcal{N}^{[j]}}{ f^{[i]} T^{[i]} }
\\ = &
    \sum_{j=1}^{n_z} P(z^{[j]}|M^{[j]})
    \sum_{i \in\mathcal{N}^{[j]}}{  - T^{[i]}\log(  P(r^{[i]}))}
\end{aligned}
\end{equation}
The computation procedure of GauSS-MI is summarized in Algorithm~\ref{alg_mutual_info}.

\begin{algorithm}[t]
\caption{GauSS-MI}
% \caption{Gaussian Splatting Shannon Mutual Information}
\label{alg_mutual_info}

\begin{algorithmic}[1]
\Require 3DGS map $\mathcal{G}$; Novel view $\bm\sigma$

    \State $I=0$
    \State $\mathcal{N} \gets \mathtt{SortGaussians} (\bm\mu, \bm\sigma)$
    \For {$j \gets  1$ \textnormal{to} $n_z$}              \Comment{Per pixel}
        \State // Rasterize color and $f$  \eqref{eq_info_gain_comp} on each pixel
        \State ${C}^{[j]} = 0$; $f^{[j]} = 0$; $T^{[j]} = 1$
        % \State $f^{[j]} = 0$
        % \State $T^{[j]} = 1$
        \For {$i \gets 1$ \textnormal{to} $\mathcal{N}^{[j]}$}  \Comment{Per gaussian}
            \State $T^{[i]} =  \alpha^{[i]}T^{[j]}$             \Comment{Gaussian's transmittance}
            \State ${C}^{[j]} \mathrel{{+}{=}} c^{[i]}T^{[i]}$                \Comment{Rasterize color}
            \State $f^{[j]} \mathrel{{-}{=}} \log(P^{[i]})T^{[i]}$            \Comment{Rasterize \eqref{eq_info_gain_comp}}
            \State $T^{[j]} =  (1-\alpha^{[i]})T^{[j]}$         \Comment{Update pixel transmittance}
        \EndFor
        \State $M^{[j]} \gets \mathtt{Color2Luminance} ({C}^{[j]})$
        \State $P(z^{[j]} | M^{[j]}) \gets \mathtt{SensorModel} (M^{[j]})$
        \State $I \mathrel{{+}{=}}P(z^{[j]} | M^{[j]}) f^{[j]}$               \Comment{Update MI}
    \EndFor
    \State \textbf{return} $I$
\end{algorithmic}
\end{algorithm}

%-----------------------------------% Section - implementation %-----------------------------------%
\section{System Implementation}
\label{sec_sys_imple}
This section details the system implementation of GauSS-MI.

\subsection{System Overview}

\begin{figure}[t] 
\centering
    \includegraphics[width=0.8\linewidth]{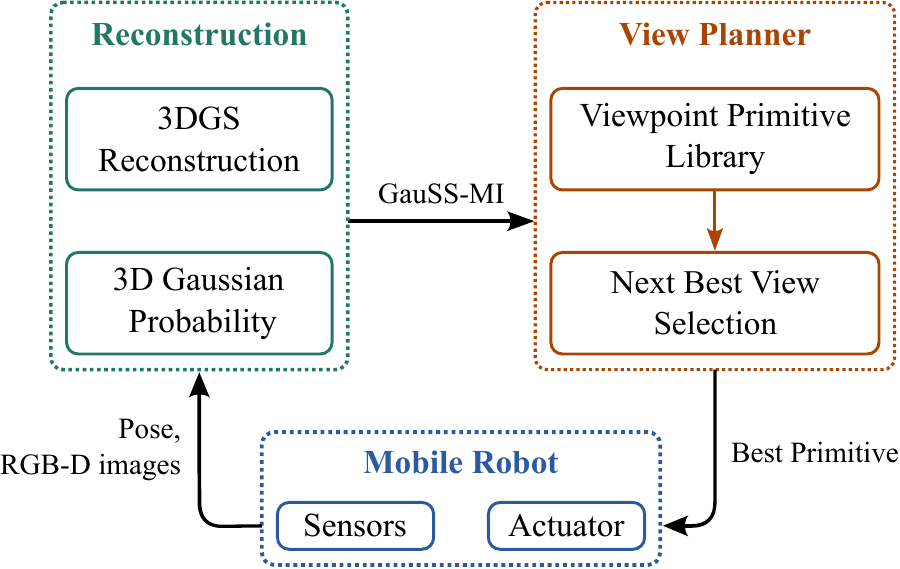}
    \caption{
        Overview of proposed active 3D reconstruction system. 
    }
    \label{fig_imple_overview}
\end{figure}

The proposed active reconstruction system comprises a reconstruction module and a planning module, as illustrated in Figure~\ref{fig_imple_overview}. 
In this work, a mobile robot is equipped with sensors that can capture color images and depth images and estimate its pose. 
Given these messages, the reconstruction module constructs and updates a 3D Gaussian splatting (3DGS) model in real-time, while simultaneously generating the 3D Gaussian probability map. 
Meanwhile, the planning module creates a library of candidate viewpoints along with the primitive trajectories.
The optimal viewpoint and primitive trajectory are subsequently determined by evaluating both the viewpoint's GauSS-MI and the trajectory's motion energy cost. 
The robot then follows the selected primitive trajectory and captures images from the next-best viewpoint.
Given the new observations, the reconstruction module could update the map.
The process iterates and results in a high-quality 3D reconstruction with detailed visual representation.

We then present the view planner for viewpoint sampling and selection and discuss the autonomous termination condition design for the proposed system.

%-----------------------------------% SubSection - Planning %-----------------------------------%
\subsection{View Planning}
\label{sec_viewplanner}

\subsubsection{Viewpoint Primitive Library}\label{sec_vpll}

% To determine the next best viewpoint, we design a viewpoint candidates generation library and choose the next best view within the library. 
To determine the next best viewpoint, we design an action library to generate a set of candidate viewpoints, and choose the next best view within the candidates. % \mathcal{A} -> \mathcal{sigma} - > the best sigma %
Inspired by the action generation method proposed in \cite{yang2017framework}, we design the action to the next viewpoint by
%each candidate viewpoint can be derived by an action
%we design the viewpoint library by sampling the action 
$$\bm \alpha=[v_{\rm xy}, v_{\rm z}, \omega_{\rm z}]$$
where $v_{\rm xy}$ and $v_{\rm z}$ represent the body frame linear velocity in ${\rm x}_\mathcal{B} - {\rm y}_\mathcal{B}$ plane and ${\rm z}_\mathcal{B}$ direction, 
%the horizontal linear velocity and vertical linear velocity in the body frame, 
and $\omega_{\rm z}$ is the body frame angular velocity around the ${\rm z}_\mathcal{B}$ axis. 
We simplify the action of 2-dimensional horizontal movement into 1 dimension, which can be compensated through the $\omega_{\rm z}$ rotation.
% The action space is given by uniformly sampled under the customized maximum velocity bounds
The action space is given by sampling each velocity that,
\begin{equation}
    \mathcal{A} = \{
        \bm \alpha| v_{\rm xy}\in \mathcal{V}_{\rm xy},  
        v_{\rm z}\in\mathcal{V}_{\rm z}, 
        \omega_{\rm z}\in{\Omega_{\rm z}} \}
\end{equation}

%% generate the viewpoint
In this paper, we assume that the sensor, normally with a limited field of view, is equipped forward, i.e., facing the ${\rm x}_\mathcal{B}$ axis. 
Thus, in the further forward propagation derivation, we design the horizontal movement action $v_{\rm xy}$ works on the ${\rm y}_\mathcal{B}$ axis. 

Given the action 
$\bm \alpha=[v_{\rm xy}, v_{\rm z}, \omega_{\rm z}] \in\mathcal{A}$
%$\bm \alpha^k=[v_{\rm xy}^k, v_{\rm z}^k, \omega_{\rm z}^k] \in\mathcal{A}$
, the next viewpoint is designed by forward propagation with duration time $T$, 
\begin{equation}\label{eq_viewpoint_i}
\begin{split}
    \bm\sigma_f &= \bm\sigma_0 +
    \begin{bmatrix}
        - v_{\rm xy} T\sin{(\psi_0+ \omega_{\rm z}  T)}
        \\ 
        v_{\rm xy}  T  \cos{(\psi_0+ \omega_{\rm z}  T)}
        \\
        v_{\rm z} T
        \\
        \omega_{\rm z}  T
    \end{bmatrix} 
\\
    \bm\sigma^{(n)}_f &= \bm 0 \quad  {\rm for}\,\, n=1,2,3
\end{split}
\end{equation}
where ${(\cdot)}^{(n)}$ denotes the $n$-th derivatives, which constraints the final state to ensure a stable picture taking on the next viewpoint.
A motion primitive trajectory $\bm\sigma_T$ from current state $\bm\sigma(t) = \bm\sigma_0$ to the next viewpoint $\bm\sigma(t+T) = \bm\sigma_f$ can be derived in closed-form \cite{mueller2015computationally} (detailed in the Appendix~\ref{apd_sec_minsnap_planner}).

Overall, by defining the set of actions $\mathcal{A}$, given the current state $\bm\sigma_0$, a library of candidate viewpoints $\bm\Sigma_f=\{\bm\sigma_f \}$ along with the primitive trajectories $\bm\Sigma_T=\{\bm\sigma_T\}$ can be formed as a viewpoint primitive library. 

\subsubsection{Next Best View Selection}
\label{sec_nbv_select}

The total reward for the next best view evaluation includes the mutual information $I$ \eqref{eq_MI_compute} and the motion cost $J$ as
\begin{equation}
    \label{eq_nbv_reward}
    R = w_II-w_JJ
\end{equation}
where $w_I, w_J>0$ are constant reward weights to balance the range of two components. 
The motion cost $J$ can be calculated based on the trajectory $\bm\sigma_T$ with respect to a specific mobile robot. 
The next best view with primitive $\bm\sigma^{*}_T$ is selected by optimizing $R$ over all feasible primitives, which is then assigned to the controller for tracking.
The complete procedure for view and motion planning is summarized in Algorithm~\ref{alg_viewplanner}.

\begin{algorithm}[t]
\caption{View and Motion Planner}
\label{alg_viewplanner}

\begin{algorithmic}[1]
\Require Full Quadrotor States $\mathcal{X}(t)$; Action Space $\mathcal{A}$

    \State $R=0$
    \For {$\bm \alpha \in \mathcal{A}$}              %\Comment{Per pixel}
        \State $\bm \sigma (t+T)\gets \mathtt{ForwardPropagate}(\bm\sigma(t)$,$ \bm \alpha)$
        \State $\bm\sigma_T \gets \mathtt{MotionPrimitive}(\mathcal{X}(t)$,$\bm \sigma (t+T))$
        
        \If{$\mathtt{SafetyCheck}(\bm\sigma_T)$}
            \State $I\gets \mathtt{GauSS\_MI} (\bm\sigma (t+T))$ \Comment{Algorithm\ref{alg_mutual_info}}
            \State $J\gets \mathtt{MotionCost} (\bm\sigma_T)$
            
            \If{$R<w_II-w_JJ$}
                \State ${R}=w_II-w_JJ$
                \State $\bm\sigma ^*_T=\bm\sigma _T$
            \EndIf
        \EndIf
    \EndFor
    \State \textbf{return} $\bm\sigma^*_T$
\end{algorithmic}
\end{algorithm}

% \subsection{Real-time System Implementation}
% we are actually predicting NBV 2 steps later. Prove that the MI computation formula does not change. 

\subsection{Termination Condition}

A spherical Gaussian may be deemed rendering reliable from one perspective, but this reliability may not hold from another perspective. 
Specifically, the \textit{reliable} probability of a Gaussian $P(r^{[i]})$ should exhibit anisotropic behavior.
In the implementation, to address this issue, we update $P(r^{[i]})$ from four orthogonal horizontal perspectives. 
The visual reconstruction completeness of a Gaussian is quantified based on the average \textit{reliable} probability denoted as $\mu_P$.
% , and actively terminate the reconstruction process based on the average \textit{reliable} probability $\mu_P$. 

At the beginning of the mapping process, we assume no prior information about the environment. 
Consequently, we initialize the probabilities of all 3D Gaussians as \eqref{eq_imple_init_prob}. 
As detailed in Section~\ref{subsec_3dgs_prob}, the probability $P(r)$ for a spherical Gaussian decreases if it renders out a relatively large image loss. 
As the reconstruction process progresses, we expect a decrease in the render-ground truth loss and an increase in $P(r)$.
A Gaussian is considered completely reconstructed when its average probability exceeds a specified threshold, $\mu_P > \tau$. 
The active reconstruction process is regarded complete and actively terminated once the proportion of fully reconstructed Gaussians reaches a predefined percentage threshold, % $\varphi$:
\begin{equation} \label{eq_termin_conditon} 
    \frac{N_{\text{done}}}{N_{\text{GS}}} > \varphi
\end{equation}
where $N_{\text{done}}$ represents the number of completely reconstructed Gaussians, and $N_{\text{GS}}$ denotes the total number of Gaussians in the map.

%-----------------------------------% Section Experiments %-----------------------------------%
\section{Simulation Experiments}

In this section, we present a series of simulation experiments designed to validate the proposed method. 
We begin by detailing the experimental setup and evaluation metrics.
Based on this, we initially validate the proposed system (Section~\ref{sec_sim_system_validate}). 
Subsequently, we conduct experiments to evaluate the proposed GauSS-MI metric from multiple perspectives: the efficiency of next-best-view selection (Section~\ref{sec_sim_viewselect}), real-time computational performance (Section~\ref{sec_sim_CompEff}), and the effectiveness of uncertainty quantification (Section~\ref{sec_sim_UncertQuanti}).
Finally, we compare the complete system against baseline methods in Section~\ref{sec_sim_activeRecon} and study the termination condition for the system in Section~\ref{sec_sim_terminate}.

%-----------------------------------% SubSection - Setup %-----------------------------------%

\subsection{System Validation}
\label{sec_sim_system_validate}
\subsubsection{Simulation Setup}
% {\color{red} may be combined and named system setup and result}

The simulation environment is created using Flightmare\cite{song2020flightmare}, featuring a configurable rendering engine within Unity\footnote{\url{https://unity.com/}} and a versatile drone dynamics simulation. 
% The simulation environment is implemented based on Flightmare\cite{song2020flightmare}, which is featured by a configurable rendering engine built upon Unity\footnote{https://unity.com/} and a flexible drone dynamics simulation. 
A quadrotor is employed as the agent for active reconstruction, equipped with an image sensor providing RGB-D images at a resolution of $640 \times 480$ and a $90\deg$ Field of View (FOV). 
% In the simulation, we use a quadrotor as the agent for active reconstruction, which is flexible for cluttered scenes.
% The quadrotor is equipped with an image sensor that provides RGB-D images with a resolution of $640 \times 480$ and a Field of View (FOV) of $90\deg$. 
The online 3D Gaussian splatting reconstruction is developed based on MonoGS\cite{Matsuki2024CVPR}, which incorporates depth measurements to enhance the online reconstruction model.
Both the proposed active reconstruction system and the simulator operate on a desktop with a 32-core i9-14900K CPU and an RTX4090 GPU.
The parameters of the proposed system are summarized in Table~\ref{tab_params}.

\begin{table}[t]
\centering
\caption{
Parameters of the Proposed System
}
\label{tab_params}
\begin{tabular}{ll}
\toprule
    \textbf{Parameter}            					& \textbf{Value}		\\ 
\midrule 
    hyperparameter on loss $\lambda_L$                      & $1.7$     \\
    hyperparameter on cumulative transmittance $\lambda_L$  & $7.0$     \\
    Primitive duration time $T$                             & \SI{1.6}{\second} \\%$1.6s$     \\
    reward weight on information $w_I$                      & $0.03$     \\
    reward weight on motion cost $w_J$                      & $0.01$     \\     
            %% motion cost is very high ~ 500~1000 per primitive
    probability threshold $\tau$                            & $0.7$     \\
    reconstruction terminate threshold $\varphi$            & $75\%$    \\
\bottomrule 
\end{tabular}
\end{table}

\subsubsection{Metrics}\label{subsec_metrics}
The evaluation focuses on assessing the visual quality of the reconstruction results and the efficiency of the active reconstruction process. 
Visual quality is evaluated using Peak Signal-to-Noise Ratio (PSNR), Structural Similarity Index (SSIM), and Learned Perceptual Image Patch Similarity (LPIPS) to quantitatively compare rendered images from the 3DGS model with a testing dataset of ground-truth images. 
Efficiency is measured by calculating the total length of the reconstruction path $P$ and the number of frames $N_f$. 
To provide a quantitative assessment of the efficiency of the reconstruction process, we introduce an efficiency metric that combines visual quality and motion effort, defined as 
$E = {\rm PSNR} / \log{N_{f}}$.
The logarithmic transformation of the denominator is applied to align with the PSNR calculation.

%-----------------------------------% SubSection - Results %-----------------------------------%

%--------------------% Result Figure %--------------------%
\begin{figure*}[t] 
    \centering
    \includegraphics[width=0.9\linewidth]{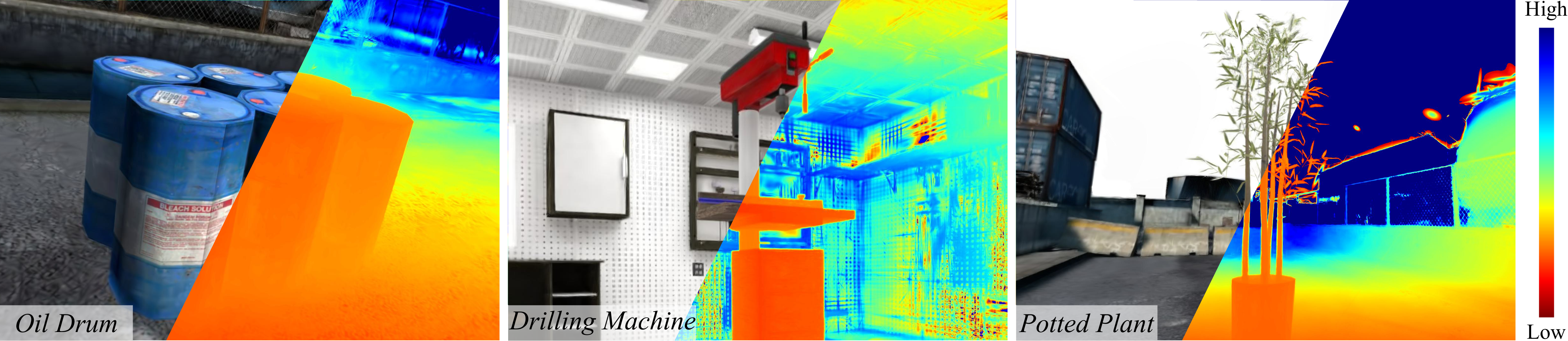}
    \caption{
        High-resolution novel view synthesis of the reconstruction result by the proposed system: color rendering against depth rendering.
    }
    \label{fig_expr_result}
\end{figure*}

%--------------------% Compare Table %--------------------%

\begin{table*}[t]
\centering
\caption{
Evaluation Results and Comparison of Simulation Experiments
% {\color{red}[more illustrate]}
}
\label{tab_expr_result}
\begin{threeparttable}
    \setlength{\tabcolsep}{2pt}
\begin{tabular}{@{}c|cccccc|cccccc|cccccc@{}}
\toprule
    Scene $^1$
    & \multicolumn{6}{c|}{ \textit{Oil Drum $^2$}} 
    & \multicolumn{6}{c|}{ \textit{Drilling Machine $^2$}} 
    & \multicolumn{6}{c} { \textit{Potted Plant $^2$}} \\ 
\midrule
    Metric 
    & \multicolumn{3}{c|}{Visual Quality} & \multicolumn{3}{c|}{Efficiency} 
    & \multicolumn{3}{c|}{Visual Quality} & \multicolumn{3}{c|}{Efficiency} 
    & \multicolumn{3}{c|}{Visual Quality} & \multicolumn{3}{c}{Efficiency} \\ 
\midrule
    Method 
    & PSNR$\uparrow$ & SSIM$\uparrow$ & \multicolumn{1}{c|}{LPIPS$\downarrow$} 
    & $N_{f}$ $\downarrow$ & $P {\rm (m)}$ $\downarrow$ & $ E \uparrow$ 
    & PSNR$\uparrow$ & SSIM$\uparrow$ & \multicolumn{1}{c|}{LPIPS$\downarrow$} 
    & $N_{f}$ $\downarrow$ & $P {\rm (m)}$ $\downarrow$ & $ E \uparrow$ 
    & PSNR$\uparrow$ & SSIM$\uparrow$ & \multicolumn{1}{c|}{LPIPS$\downarrow$} 
    & $N_{f}$ $\downarrow$ & $P {\rm (m)}$ $\downarrow$ & $ E \uparrow$ \\ 
\midrule
    Ours 
    & \textbf{34.35} & \textbf{0.986} & \multicolumn{1}{c|}{ \textbf{0.068} } & \textbf{141} & 61.04 & \textbf{16.0}
    & \textbf{33.99} & \textbf{0.995} & \multicolumn{1}{c|}{ \textbf{0.040} } & \textbf{122} & 36.16 & \textbf{16.3}
    & 30.33 & 0.986 & \multicolumn{1}{c|}{ 0.084 } & \textbf{200} & 79.60 & \textbf{13.2}
    \\
    FUEL~\cite{zhou2021fuel}
    & 22.82 & 0.915 & \multicolumn{1}{c|}{ 0.186 } & 165 & \textbf{15.21} &  10.3
    & 21.08 & 0.967 & \multicolumn{1}{c|}{ 0.116 } & 145 & \textbf{11.16} &  9.8
    & 25.39 & 0.963 & \multicolumn{1}{c|}{ 0.149 } & {205} & \textbf{17.28} & 11.0
    \\
    NARUTO~\cite{feng2024naruto}
    & 31.84 & 0.976 & \multicolumn{1}{c|}{ 0.072 } & 3000 & 116.34 & 9.2
    & 31.50 & 0.992 & \multicolumn{1}{c|}{ 0.047 } & 3000 & 92.35  & 9.1
    & \textbf{30.83} & \textbf{0.988} & \multicolumn{1}{c|}{ \textbf{0.057} } & 4000 & 157.75 & 8.6
    \\ 
\bottomrule
\end{tabular}

\begin{tablenotes}
    \item[1] Simulation scenes are built by Flightmare~\cite{song2020flightmare}. 
    \item[2] \textit{Oil drum} scene size: $\rm 5 m\times4 m\times3 m$. 
    \textit{Drilling Machine} scene size: $\rm 4 m\times4 m\times3 m$. 
    \textit{Potted Plant} scene size: $\rm 5 m\times5 m\times5 m$.
\end{tablenotes}

\end{threeparttable}
\end{table*}

%-----------------------------% Compare Table %-----------------------------%

\subsubsection{Simulation Result}

We actively reconstruct three scenes, the \textit{Oil Drum}, the \textit{Drilling Machine}, and the \textit{Potted Plant}, to validate the proposed system.
The offline refinement results, including image rendering and depth rendering, are presented in Figure~\ref{fig_expr_result}. 
The evaluations of visual quality and efficiency are calculated and summarized in Table~\ref{tab_expr_result}. 
The \textit{Oil Drum} is characterized by a relatively simple geometry but detailed texture. 
The \textit{Drilling Machine} exhibits fine geometric structures, while the \textit{Potted Plant} features a highly cluttered geometric structure. 
The rendering results in Figure~\ref{fig_expr_result} demonstrate a detailed visual fidelity with precise geometric structures, highlighting the system's capability to capture intricate textures and structures.

%-----------------------------------% SubSection - active view %-----------------------------------%

\subsection{Comparison Study of Active View Selection}
\label{sec_sim_viewselect}

To evaluate the efficiency of the proposed GauSS-MI metric in selecting the next-best-view for high visual quality reconstruction, we conduct comparative experiments on active view selection using a fixed number of frames.

\subsubsection{Baselines}
We benchmark our method against FisherRF\cite{jiang2025fisherrf}\footnote{FisherRF: \url{https://github.com/JiangWenPL/FisherRF}}, a state-of-the-art radiance field-based active view selection approach that quantifies the expected information gain by constructing the Fisher information matrix. 
To ensure a fair comparison, FisherRF is integrated into our system by substituting the GauSS-MI evaluation $I$ in \eqref{eq_nbv_reward} with its FisherRF metric.
%\eqref{eq_MI_compute}.
Additionally, a random view selection policy is implemented as a baseline to highlight the benefits of using view selection strategies.

\subsubsection{Results}
\begin{figure}[t] 
    \centering
    \includegraphics[width=0.85\linewidth]{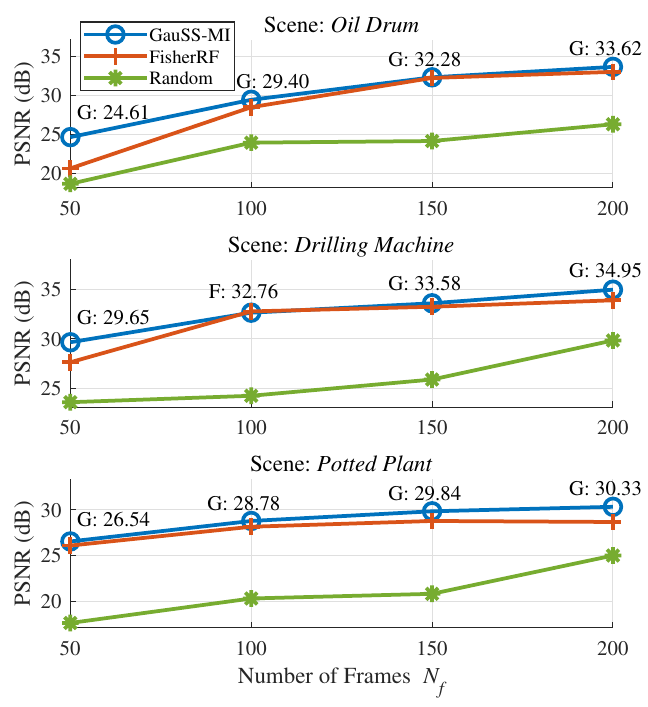}
    \caption{PSNR results for active view selection with a limited number of frames.
    The maximum PSNR value for each test is annotated. The abbreviations 'G' and 'F' denote GauSS-MI and FisherRF, respectively.
    % {\color{red}[add scene name in figure title]}
    }
    \label{fig_view_select_psnr}
\end{figure}

\begin{figure*}[t] 
    \centering
    \includegraphics[width=0.95\linewidth]{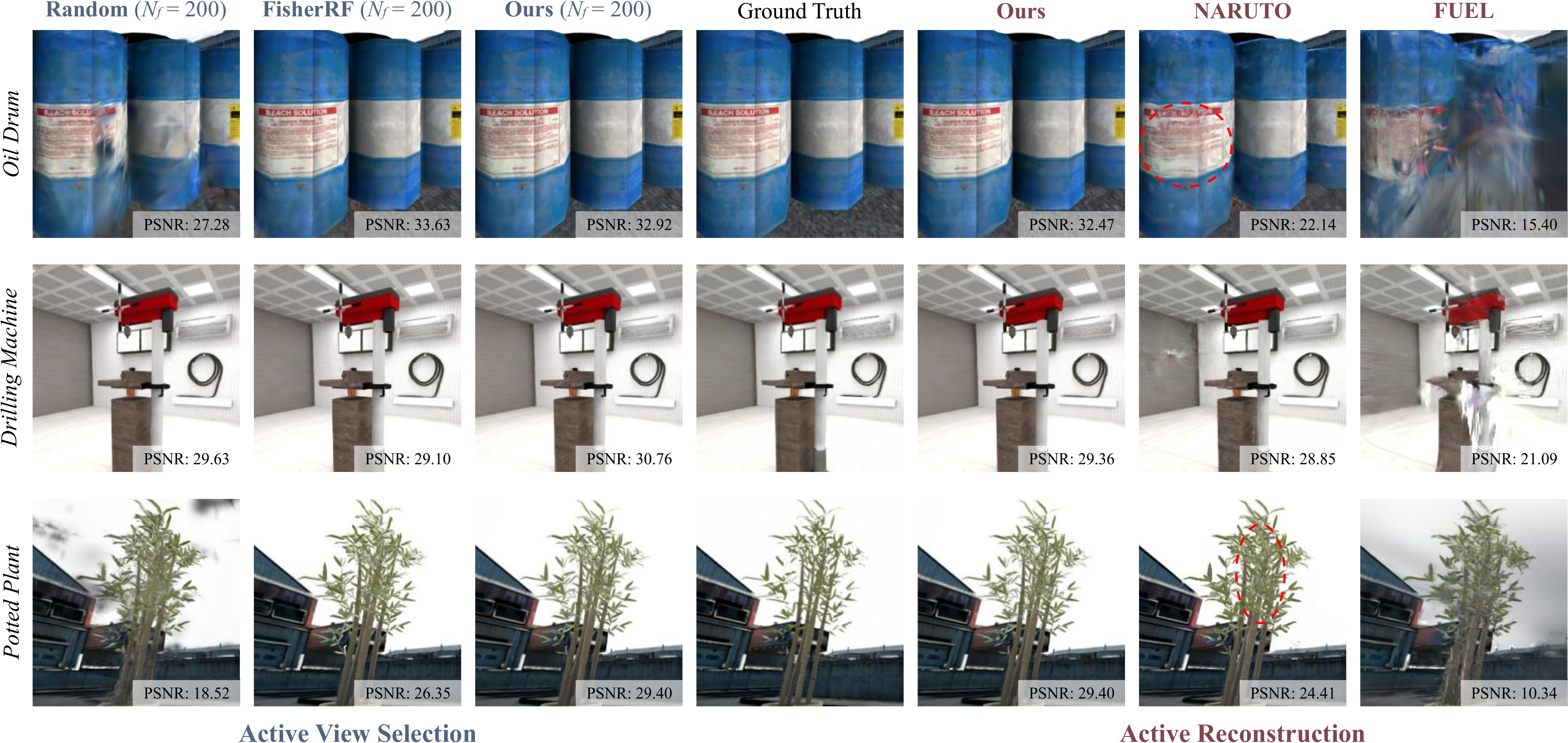}
    \caption{
Novel view synthesis results compared to ground truth.
\textbf{Left part}: Results of active view selection with a fixed number of frames $N_f=200$. \\
\textbf{Right part}: Results of active reconstruction, with number of frames $N_f$ specified in Table~\ref{tab_expr_result}.
    }
    \label{fig_compare_novel_view}
\end{figure*}

The comparative experiment is performed across three scenes, with the number of frames limited and gradually increased for each method. 
We compute the PSNR values for each test and visualize the results by plots in Figure~\ref{fig_view_select_psnr}.
The results show that both GauSS-MI and FisherRF significantly outperform the random policy, demonstrating the methods' effectiveness in next-best-view selection for enhancing visual quality.
While the performance of GauSS-MI and FisherRF is comparable, GauSS-MI achieves higher PSNR values in most tests, validating its superior efficiency in active view selection. 
% The results of GauSS-MI and FisherRF are quite close, while GauSS-MI surpasses in most tests, validating the higher efficiency of the proposed method.
The novel view synthesis results for GauSS-MI, FisherRF, and the random policy, with a fixed number of frames $N_f=200$, are presented alongside the ground truth on the left-hand side of Figure~\ref{fig_compare_novel_view}.
These visualizations further showcase the enhanced visual fidelity reconstruction result of GauSS-MI, particularly in scenes featuring complex geometric or textural details. 
The efficiency on active view selection is especially advantageous for onboard active reconstruction, where constrained computational and battery resources necessitate minimizing the number of frames and reconstruction time.
% This active view selection efficiency is particularly valuable for onboard active reconstruction, where the onboard resources are constrained, and reducing the number of frames and reconstruction time is critical.

%-----------------------------------% SubSection - Computation Analysis %-----------------------------------%

\begin{figure}[t]
  \centering
  \includegraphics[width=0.8\linewidth]{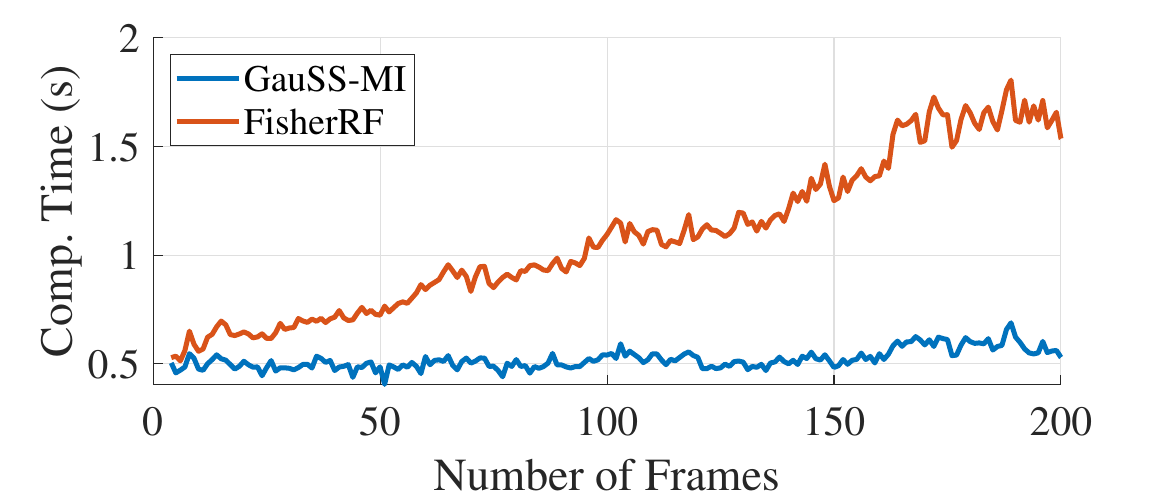}
  \caption{Comparison of computation time. Statistics in a complete active reconstruction process.}
  \label{fig_runtime}
\end{figure}

\subsection{Computational Efficiency}
\label{sec_sim_CompEff}
We analyze the computational complexity of the proposed GauSS-MI method, measure its average runtime, and compare it with FisherRF~\cite{jiang2025fisherrf}, validating the real-time performance of our metric.

\subsubsection{Computational Complexity}
The computation of GauSS-MI, as detailed in Algorithm~\ref{alg_mutual_info}, is similar to 3DGS rasterization in that Eq.~\eqref{eq_MI_compute} projects the information gain function~\eqref{eq_info_gain_comp} onto an image.
The algorithm is implemented in parallel using CUDA. 
Assuming the current 3DGS map with $N_g$ Gaussians, the image with $N_p$ pixels, and $N_c$ candidate viewpoints to be evaluated, the computational complexity of GauSS-MI is $O(N_pN_gN_c)$.
In contrast, FisherRF's complexity depends on both candidate and observed views. 
% For FisherRF, the computation is also related to the number of observed views.
With $N_o$ observed views, FisherRF requires a complexity of  $O( N_pN_g (N_o+N_c) )$ to evaluate all candidates, as it has to compute the information from both observed and candidate views at each decision step.
Consequently, the computational cost scales linearly with $N_o+N_c$, indicating the increasing runtime as active reconstruction progresses.
GauSS-MI, however, maintains consistent computation, scaling linearly with only $N_c$. 
% In contrast, GauSS-MI maintains a consistent computational load that is only linear to $N_c$. 
This efficiency stems from our probabilistic model, which quantifies the information from prior observations with a low computational overhead of $O( 2N_{p}N_g)$ during the map update process (Algorithm~\ref{alg_prob_update}). 
As a result, in the next-best-view decision step, GauSS-MI evaluates only candidate views, achieving low and stable computational complexity, making it ideal for real-time applications.

\subsubsection{Runtime} 
We conducted a complete active reconstruction experiment to measure the runtime of each method at each planning timestep, as shown in Figure~\ref{fig_runtime}. 
% GauSS-MI achieves an average runtime of $5.55~{\rm ms}$ ($182.2~{\rm fps}$), while FisherRF averages $11.66~{\rm ms}$ ($85.8~{\rm fps}$).
GauSS-MI achieves an average runtime of \SI{5.55}{\milli\second} (\SI{182.2}{\fps}), while FisherRF averages \SI{11.66}{\milli\second} (\SI{85.8}{\fps}).
% frame rate of 182.2~fps, while FisherRF averages 85.8~fps.
These results corroborate the computational complexity analysis that GauSS-MI maintains consistent runtime throughout the reconstruction process, whereas FisherRF's runtime increases due to its dependence on the growing number of observed views.

%-----------------------------------% SubSection - Uncertainty Quantification %-----------------------------------%

\subsection{Uncertainty Quantification}
\label{sec_sim_UncertQuanti}
To evaluate the uncertainty quantification capability of the proposed method, we employ sparsification plots and the Area Under Sparsification Error (AUSE) metrics~\cite{ilg2018uncertainty, goli2024bayes} to evaluate and compare our method with the state-of-the-art 3DGS-based uncertainty quantification method, FisherRF~\cite{jiang2025fisherrf}.
% , across multiple scenes.

\subsubsection{Sparsification Plots}
\begin{figure}[t] 
    \centering
    \includegraphics[width=0.8\linewidth]{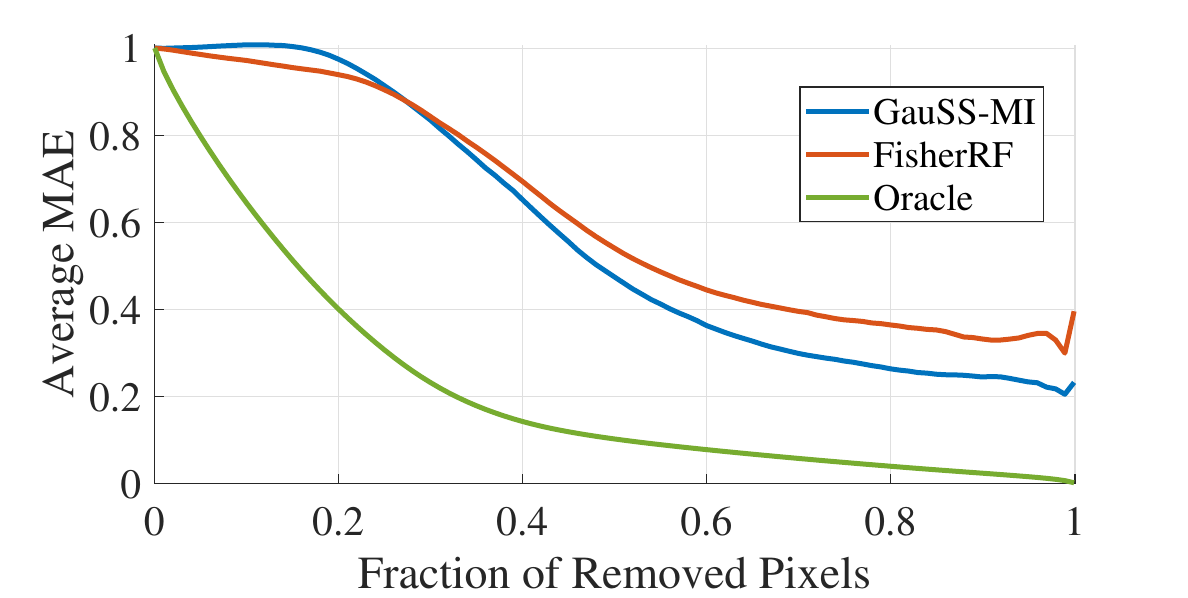}
\caption{Sparsification plots. 
The plot shows the mean absolute error (MAE) of the true error image against the fraction of pixels with the highest uncertainties removed. 
The oracle sparsification represents the lower bound, derived by removing pixels ranked by ground-truth error. 
The sparsification plots reveal the correlation between the estimated uncertainty and the true errors.
}
    \label{fig_ause}
\end{figure}

Sparsification plots provide a measurement of the correlation between the estimated uncertainty and the true errors~\cite{ilg2018uncertainty}. 
If the estimated uncertainty accurately reflects model uncertainty, progressively removing pixels with the highest uncertainty should lead to a monotonic decrease in the mean absolute error (MAE) of the true error image. 
The plot of the MAE against the fraction of removed pixels is called \textit{Sparsification Plots}, as shown in Figure~\ref{fig_ause}.
The ideal uncertainty ranking can be obtained by ordering pixels according to their true error relative to the ground truth, yielding the \textit{Oracle Sparsification} curve.
We evaluate the uncertainty estimates for all images in the test dataset across three scenes and compute the average sparsification plot.
We compare the result with FisherRF in Figure~\ref{fig_ause}. 
The plot reveals that our uncertainty estimate is closer to this oracle, indicating a stronger correlation between our predicted uncertainties and actual errors.

\subsubsection{Area Under Sparsification Error (AUSE)}
To quantitatively assess the divergence between the sparsification plot and the oracle, we calculate the Area Under Sparsification Error (AUSE)~\cite{ilg2018uncertainty}, which measures the area between the two curves.
The AUSE values for each scene are reported in Table~\ref{tab_ause}. 
Our method consistently achieves lower AUSE scores compared to FisherRF, demonstrating its superior uncertainty quantification performance.

\begin{table}[t]
\centering
\caption{
Area Under Sparsification Error (AUSE) Results
}
\label{tab_ause}
\begin{tabular}{@{}c|ccc@{}}
\toprule
    Scene
    & \textit{Oil Drum}
    & \textit{Drilling Machine}     %& \textit{Drill. Mach.}
    & \textit{Potted Plant}       %& \textit{Pot. Plant}
    % & average
    \\ 
\midrule
    Ours 
    & 0.264 & 0.498 & 0.351 %& 0.371
    \\
    FisherRF~\cite{jiang2025fisherrf}
    & 0.276 & 0.605 & 0.392  %& 0.424
    \\
\bottomrule
\end{tabular}

\end{table}

% \subsection{Ablation Study}

%-----------------------------------% Compare study %-----------------------------------%
% \subsection{Comparison Study}
\subsection{Comparison Study of Active Reconstruction}
\label{sec_sim_activeRecon}
This section evaluates and compares the complete system, including the proposed view planning and active termination condition. %, against baseline methods.
We select the state-of-the-art baselines employing different map representations and uncertainty quantification techniques to validate our system's efficiency on visual quality.

\subsubsection{Baselines} 
To evaluate the efficacy of our proposed method, we conducted a comparative analysis between our active reconstruction system and existing systems, FUEL~\cite{zhou2021fuel} and NARUTO~\cite{feng2024naruto}. 
FUEL is a volumetric-based active reconstruction system with no consideration of visual quality, while NARUTO is a NeRF-based framework that addresses radiance field uncertainty with a focus on geometry.
For our study, we implemented the comparison using the open-source codes for FUEL\footnote{FUEL: \url{https://github.com/HKUST-Aerial-Robotics/FUEL}} and NARUTO\footnote{NARUTO: \url{https://github.com/oppo-us-research/NARUTO}}, employing their default parameter settings. 
Each system, including the next best view selection and path planning algorithm, captured color images in the three simulation scenes, which are subsequently employed to 3D Gaussian Splatting~\cite{kerbl20233d} for offline model reconstruction. 
Evaluation of reconstruction quality and efficiency was conducted using the metrics outlined in Section~\ref{subsec_metrics}.

\subsubsection{Results} 

% overall
The quantitative results are presented in Table~\ref{tab_expr_result}, while the qualitative visual comparisons are shown in the right part of Figure~\ref{fig_compare_novel_view}. 
Our system demonstrates superior efficiency across all scenes and attains the highest visual quality in the \textit{Oil Drum} and \textit{Drilling Machine}. 
% 1 NARUTO better in s3
In the \textit{Potted Plant} scene, NARUTO slightly outperforms our system by a small margin. 
However, it is worth noting that NARUTO completed its reconstruction process after capturing thousands of images, which contributed to its commendable reconstruction performance. 
The extensive collection of images is attributed to NARUTO's continuous high-frequency image capture throughout its movement.
The abundance of images with significant overlap resulted in a lower active viewpoint selection efficiency, indicating an inadequate assessment of observed information and a suboptimal reconstruction strategy.
In contrast, our system efficiently selects viewpoints guided by GauSS-MI. 
As a result, we achieve comparable or even superior visual quality to NARUTO while maintaining consistently high efficiency.

% 2 FUEL better for path
In terms of total path length for active reconstruction, FUEL stands out for completing the process with a notably shorter trajectory compared to both our system and NARUTO. 
This outcome aligns with expectations, given that FUEL focuses solely on geometric completeness during active reconstruction. 
However, despite its efficiency in path length, FUEL consistently yields the lowest visual quality and the reconstruction results exhibit poor texture quality, as illustrated in Figure~\ref{fig_compare_novel_view}. 
This indicates the inadequacy of relying solely on geometric evaluation for high-quality visual reconstructions.

% conclusion
Overall, our system excels in active efficiency while simultaneously delivering high visual quality across all scenes. 
This demonstrates the effectiveness of our probabilistic model in evaluating observed information and the capability of GauSS-MI in identifying optimal viewpoints to enhance efficiency.

%-----------------------------------% Terminate study %-----------------------------------%

\subsection{Termination Study} 
\label{sec_sim_terminate}

\begin{figure}[t] 
    \centering
    \includegraphics[width=\linewidth]{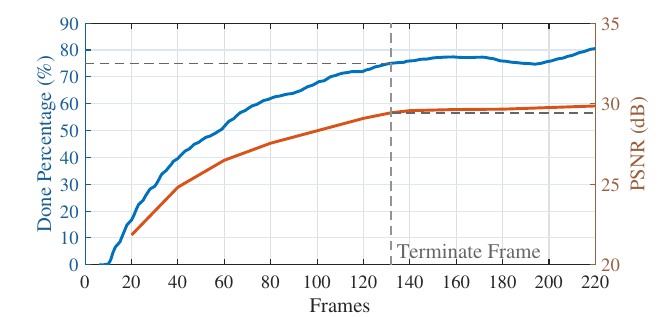}
    \caption{Terminate Condition Validation
    }
    \label{fig_terminate_test}
\end{figure}

We additionally discuss our termination condition in the active reconstruction process. We statistic the PSNR and the percentage of reconstructed Gaussians along the active reconstruction process in Figure~\ref{fig_terminate_test}, where the “Done Percentage” refers to $N_{\text{done}}/N_{\text{GS}}$ in \eqref{eq_termin_conditon} and the PSNR is calculated with offline refinement.
The figure shows that %the “Done Percentage” converges around $75\%$. 
once 75\% of the Gaussians in the map are fully reconstructed, the PSNR reaches 29.44, with minimal further improvement even as the reconstruction process continues. 
This may result from the limitations of the online 3DGS reconstruction algorithm, which is hard to achieve higher visual quality under the constraints of real-time mapping.

%-----------------------------------% Real-World Experiment %-----------------------------------%
\section{Real-World Experiments} 
% This section presents the real-world implementation and the results.

\begin{figure}[t]
    \centering
    \includegraphics[width=1.0\linewidth]{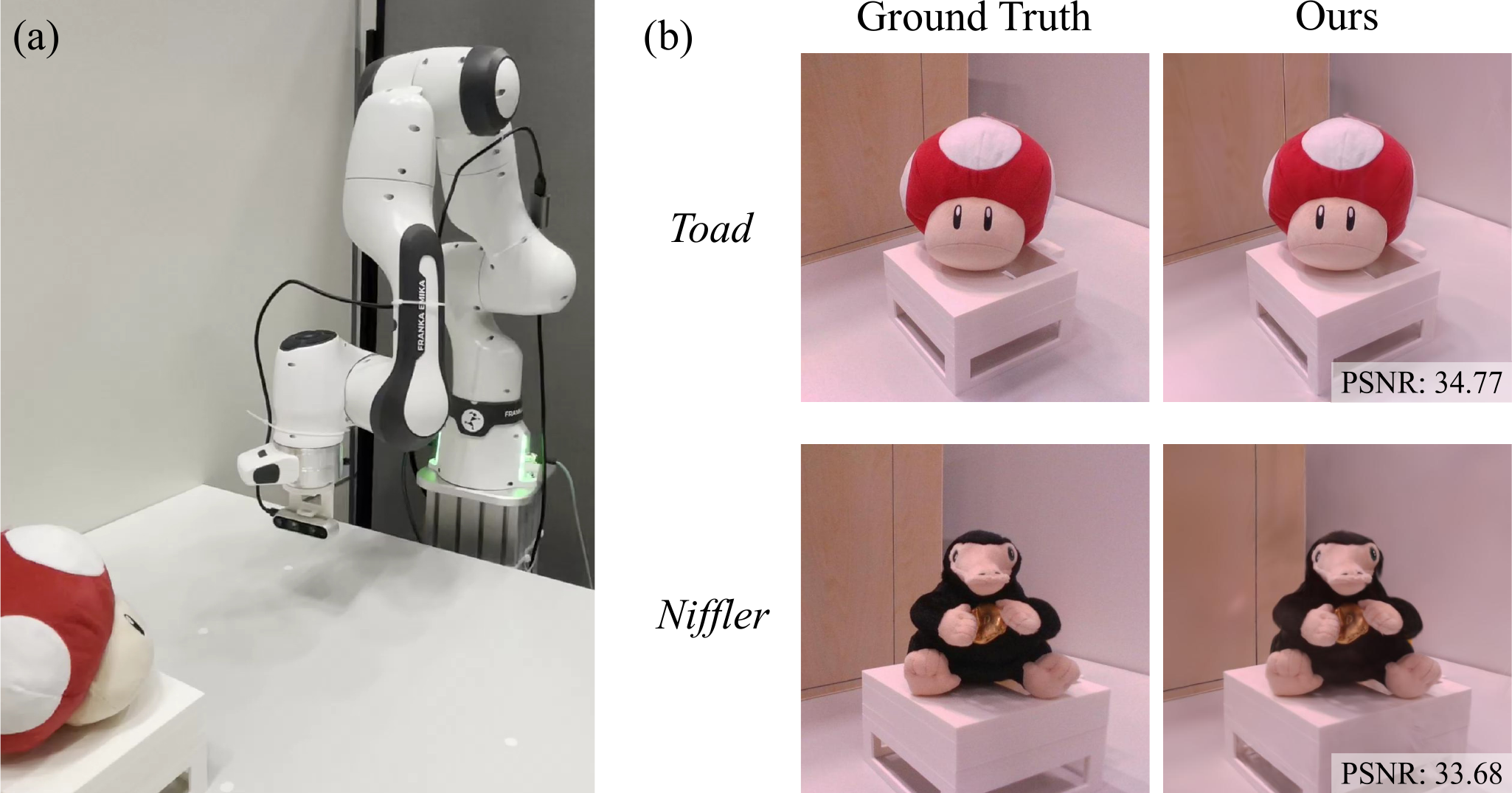}
    \caption{
    Active reconstruction experiment with GauSS-MI in the real world.
    (a) Experiment setup. (b) Novel view synthesis results.
    % {\color{red}[more illustrate]}
    }
    \label{fig_realworld_result}
\end{figure}

% \subsection{Implementation Details} 
To validate the efficacy of the proposed method in practical settings, we conduct real-world experiments using a Franka Emika Panda robotic arm equipped with an Intel RealSense D435 depth camera for capturing RGB-D images. The real-world setup is shown in Figure~\ref{fig_realworld_result}(a). 
The active reconstruction system for real-world implementation integrates the online 3DGS reconstruction algorithm with the proposed active view sampling and selection method.
% The experimental system integrated online 3DGS reconstruction algorithm with the proposed active view sampling and selection algorithm. 
Motion planning and control for the Franka arm are implemented using the MoveIt framework \footnote{\url{https://github.com/moveit/moveit}}, facilitating precise pose control and reliable feedback.
All algorithms are executed on a desktop equipped with a 32-core Intel i9-13900K CPU and an NVIDIA RTX 4090 GPU.

For the real-world demonstration, we actively reconstructed two scenes: the \textit{Toad} and the \textit{Niffler}. 
% We actively reconstruct two scenes, the \textit{Toad} and the \textit{Niffler}, for real-world demonstration. 
The \textit{Toad} scene features relatively smooth surfaces, whereas the \textit{Niffler} exhibits intricate geometric details.
Constrained by the robot arm's workspace and the minimal detection range of the depth camera, the experimental scenes are limited to the size of \SI{0.2}{\meter}$\times$\SI{0.2}{\meter}$\times$\SI{0.2}{\meter}.%$0.2{\rm m}\times0.2{\rm m}\times0.2{\rm m}$ 
The novel view synthesis results, presented in Figure~\ref{fig_realworld_result}(b), demonstrate the high visual fidelity achieved by our method.
Quantitative evaluation of visual quality and reconstruction efficiency is summarized in Table~\ref{tab_rw_expr_result}.
The results demonstrate the effectiveness and efficiency of the proposed system in the real world, highlighting its robustness across different scene complexities.

%-----------------------------% Real-World Exper Figure %-----------------------------%

%-----------------------------% Real-World Exper Table %-----------------------------%

\begin{table}[t]
\centering
\caption{
Evaluation Results of Real-world Experiments
}
\label{tab_rw_expr_result}
\begin{tabular}{c|cccccc}
% \begin{tabular}{@{}c|cccccc@{}}
\toprule
    Metric 
    & \multicolumn{3}{c|}{Visual Quality} & \multicolumn{2}{c}{Efficiency} 
    \\ 
\midrule
    Scene% $^1$
    & PSNR$\uparrow$ & SSIM$\uparrow$ & \multicolumn{1}{c|}{LPIPS$\downarrow$} 
    & $N_{f}$ $\downarrow$ & $ E \uparrow$ 
    \\ 
\midrule
    \textit{Toad}
    & 32.53 & 0.9336 & \multicolumn{1}{c|}{ 0.2693 } & 24  & 23.6
    \\ 
    \textit{Niffler}
    & 28.20 & 0.9273 & \multicolumn{1}{c|}{ 0.3020 } & 36  & 18.1
    \\ 
\bottomrule
\end{tabular}
\vspace{-0.3cm}

\end{table}

%-----------------------------------% Limitation %-----------------------------------%

\section{Limitations}\label{sec_limitation}
In this section, we discuss the limitations of our study. 

%% rely highly on online 3DGS algorithm
Firstly, the performance of our active reconstruction system is closely tied to the rendering quality of the online 3DGS framework. 
Due to real-time computation constraints, the online 3DGS algorithm is hard to match the high rendering quality achievable in offline settings. 
Future research could focus on enhancing real-time 3DGS reconstruction algorithms for active systems, potentially leading to significant improvements in reconstruction results.

Furthermore, the determination of the termination threshold $\varphi$ for achieving high visual quality was made heuristically in our study, without a systematic investigation. 
A more thorough analysis of how the termination condition, scene complexity, and reconstruction fidelity are interrelated would be valuable for refining termination criteria in future studies.

Lastly, although we have implemented the proposed algorithm on an agile and versatile quadrotor, the viewpoints are currently limited to yaw variations (rotation on the vertical $z$ body axis) only. 
Roll and pitch movements have not been taken into account. 
The whole-body movements requires comprehensive planning on the observation viewpoints and the quadrotor motion.

%-----------------------------------% Conclusion %-----------------------------------%

\section{Conclusion and Future Work}
This paper addresses a critical challenge in active reconstruction—active view selection—with a focus on enhancing visual quality. 
We first introduce an explicit probabilistic model to quantify the uncertainty of visual quality, leveraging 3D Gaussian Splatting as the underlying representation. 
Building on this, we propose Gaussian Splatting Shannon Mutual Information (GauSS-MI), a novel algorithm for real-time assessment of mutual information between measurements from a novel viewpoint and the existing map. 
GauSS-MI is employed to facilitate the active selection of the next best viewpoints and is integrated into an active reconstruction system to evaluate its effectiveness in achieving high visual fidelity in 3D reconstruction. 
Extensive experiments across various simulated environments and real-world scenes demonstrate the system’s ability to deliver superior visual quality over state-of-the-art methods, validating the effectiveness of the proposed approach.

Future research will first focus on addressing the limitations outlined in Section~\ref{sec_limitation}. Beyond this, we aim to extend our work from simulation to real-world deployment on drones. This transition introduces additional challenges, such as constrained onboard computational resources.

\section*{Acknowledgement}
This research is partially supported by the Innovation and Technology Commission of the HKSAR Government under the InnoHK initiative, Hong Kong Research Grants Council under NSFC/RGC Collaborative Research Scheme (CRS\_HKU703/24) and Joint Research Scheme (N\_HKU705/24).
Jia Pan is the corresponding author. 
Yixi Cai is the project leader.
The authors gratefully acknowledge Ruixing Jia and Rundong Li for their insightful and valuable discussions. 
We also thank the anonymous reviewers for their constructive and thoughtful feedback, which greatly enhanced this manuscript.

%-------------------------------------------------------------------------------%

%\newpage
\bibliographystyle{plainnat}
\bibliography{main.bib}

% \clearpage

%-----------------------------------% supplementary %-----------------------------------%

\appendix 

\subsection {Information Gain Function Derivation}
\label{apn_sec_infogainf}

The information gain function $f(\delta(Z), o_{1:k-1})$ is defined by \eqref{eq_MI_definition} that, 
\begin{equation}
\begin{aligned}
\label{eq_apn_info_gain_define} 
f(\delta&(Z), o_{1:k-1}) = \\
    &P(r|z_k=Z,Z_{1:k-1} )
    \log(\frac{ P(r|z_k=Z,Z_{1:k-1}) }
                { P(r |Z_{1:k-1})  } )
\end{aligned}
\end{equation}
Based on the definition of $o_{1:k-1}$ in \eqref{eq_pol_define}, we have
\begin{equation}\label{eq_apn_pzk_1}
P(r |Z_{1:k-1}) = \frac{ 1+o_{1:k-1} }
                { o_{1:k-1}  }
\end{equation}
Substitute the probability update \eqref{eq_prob_update} into \eqref{eq_apn_pzk_1}, we can derive
\begin{equation}\label{eq_apn_pzk}
P(r |z_k=Z, Z_{1:k-1}) = \frac{ 1+\delta(Z)o_{1:k-1} }
                { \delta(Z)o_{1:k-1}  }
\end{equation}
Substitute \eqref{eq_apn_pzk_1} and \eqref{eq_apn_pzk} into \eqref{eq_apn_info_gain_define},
the information gain function can be derived and simplified as,
\begin{equation}    \label{eq_apn_reward_func_f}
    f(\delta, o) = 
    \frac{ o }{o+\delta^{-1}} 
    \log(\frac{o+1}{o+\delta^{-1}}) 
\end{equation}

\subsection {Closed-form Minimum-snap motion primitive}
\label{apd_sec_minsnap_planner}

Here we present the detailed motion planner used in our system, associated with the motion cost derivation. 
The planner is designed for a quadrotor, which is agile and versatile in cluttered environments. 
The main notations of this section are listed in Table~\ref{apd_tab_notation_plan}.  

%-----------------------------% Mtion Plan Table %-----------------------------%

\begin{table}[t]
\centering
\caption{Main Notations for Motion Planning}
\label{apd_tab_notation_plan}
\begin{tabular}{@{}ll@{}}
\toprule
    Notations & Explanation \\ 
\midrule
    $\mathcal{X}$         & Full quadrotor state. \\
    $\mathcal{B}$           & The quadrotor body frame. \\
    % $\mathcal{W}$           & The world frame. \\
    $t, T$             & Current time and trajectory duration. \\
    $\bm\sigma$             & Flat outputs of a quadrotor, a viewpoint. \\
    %% %% %%
    $\bm p$             & Position in the world frame. \\
    $\bm v:=\dot{\bm p}$              & Linear velocity in the world frame. \\
    $\bm a:= \bm p^{(2)}$             & Linear acceleration in the world frame. \\
    $\bm j:= \bm p^{(3)}$             & Jerk. \\
    $\bm s:= \bm p^{(4)}$             & Snap. \\
    $\psi$             & Yaw, rotation around $\rm z_\mathcal{B}$ axis . \\
    % $\bm \omega$             & Angular velocity in the body frame. \\
    $\alpha, \beta, \gamma, \delta$             & Primitive trajectory coefficients. \\
    %% %% %%
    ${(\cdot)}^*$           & Optimal state. \\
    ${(\cdot)}_0$           & Initial state, current state. \\
    ${(\cdot)}_f$           & Final state, objective state. \\
    ${(\cdot)}_T$           & A state transition with duration $T$. \\
    %% %% %%
\bottomrule
\end{tabular}
\end{table}

%-----------------------------% Mtion Plan Table %-----------------------------%

%% any flat output can be followed, 
As quadrotor dynamics has been demonstrated differentially flat \cite{mellinger2011minimum}, any smooth trajectory with physically bounded derivatives in the space of flat outputs can be followed by the under-actuated quadrotor. 
%Considering the aim of the planning viewpoint,
Aiming at view planning, we choose the flat outputs as $\bm\sigma= [x, y, z, \psi]^T$, which includes the drone's position $\bm p=(x, y, z)$ and yaw $\psi$. 
Then, the full quadrotor states $\mathcal{X}$ can be expressed by the algebraic functions of flat outputs and their derivatives that
% $\mathcal{X}_f=[p_f,v_f,a_f,j_f]^T$
\begin{equation}\label{eq_quad_full_state}
	\mathcal{X} %% the \psi, \dot\psi here maybe confusing
		=  [\bm p, {\bm v}, {\bm a}, {\bm j}, \psi, \dot\psi]
		%:= [\bm p, \dot{\bm p}, \bm p^{(2)}, \bm p^{(3)}, \psi, \dot\psi] 
\end{equation}
where the velocity $\bm v$, acceleration $\bm a$, jerk $\bm j$ are derivatives of position $\bm p$.

As the control input, thrust and torque, can be formulated by at-most the fourth derivatives, we design trajectories that minimize the snap $\bm s$, %${\bm s}:={\bm p}^{(4)}$,
\begin{equation}
\begin{aligned}
    \label{apd_eq_min_snap_problem}
    \min \quad 
    &\frac{1}{T}\int^T_0||{\bm s(t)}||^2 dt\\
    \textrm{s.t.} \quad
    &\mathcal{X}(t) = \mathcal{X}_0\\
    &\mathcal{X}(t+T) = \mathcal{X}_f
\end{aligned}
\end{equation}
where $\mathcal{X}_0$ and $\mathcal{X}_f$ represent the initial state and final state respectively. %, $\mathcal{X}(t)$ and $\mathcal{X}(t+T)$ denotes the current state and objective state with duration time $T$. 

%% trajectory generation
To efficiently generate the trajectory from current state to the next viewpoint, %we employ the closed-form motion primitives\cite{mueller2015computationally}. 
we extend the result of the work \cite{mueller2015computationally} and derive the closed-form minimum-snap primitives. 
The optimal trajectory generation problem \eqref{apd_eq_min_snap_problem} can be decoupled into three orthogonal axes that,
% The cost of each primitive in \eqref{apd_eq_min_snap_problem} can be decoupled into three orthogonal axes.
\begin{equation}
    \label{apd_eq_motion_cost}
    J = \frac{1}{T}\int^T_0||{\bm s(t)}||^2 dt 
    = \sum_{k\in\{x,y,z\}} \frac{1}{T}\int^T_0s_k^2(t) dt
\end{equation}
The axis subscript $k$ for vectors will be simplified for the remainder of this section.
Employing Pontryagin’s minimum principle, the optimal control input for each axis can be solved for,
\begin{equation}\label{apd_eq_min_s}
	{ s}^*(t)=%\frac{1}{2}{\eta_1} t^3 +\frac{3}{2}{\eta_2} t^2 + 3{\eta_3} t +3{\eta_4}
    \frac{1}{2}\alpha t^3 +\frac{3}{2}\beta t^2 + 3\gamma t +3\delta
\end{equation}
from which the optimal position for each axis follows from integration,
\begin{equation}\label{apd_eq_prim_p}
\begin{split}
    p^*(t)= &\frac{1}{1680}\alpha t^7 +\frac{1}{240}\beta t^6 + \frac{1}{40}\gamma t^5 +\frac{1}{8}\delta t^4 \\&+\frac{1}{6}j_0t^3+\frac{1}{2}a_0t^2+v_0t+p_0\\
\end{split}
\end{equation}
% where velocity $ v:=\dot{ p}$, acceleration $ a :=  p^{(2)}$, jerk $ j:= p^{(3)}$ are derivatives of position $ p$; the subscript $0$ refers to the starting state. 
The minimum cost for each axis can be derived by substituting optimal snap \eqref{apd_eq_min_s} into \eqref{apd_eq_min_snap_problem},
\begin{equation}
\label{eq_cost}
\begin{split}
    J_k =& \frac{1}{28}\alpha^2T^6+\frac{1}{4}\alpha\beta T^5 +(\frac{9}{20}\beta^2+\frac{3}{5}\alpha\gamma)T^4 \\
    &+(\frac{3}{4}\alpha\delta+\frac{9}{4}\beta\gamma)T^3 +(3\gamma^2+3\beta\delta)T^2 
    \\&+9\gamma\delta T +9\delta^2
\end{split}
\end{equation}
The constant coefficients $\alpha$, $\beta$, $\gamma$ and $\delta$ can be expressed by algebraic functions of initial state $\mathcal{X}_0$ and final state $\mathcal{X}_f$ as
\begin{equation}\label{eq_apn_abcd_1111}
\setlength{\arraycolsep}{1.0pt}
    \begin{bmatrix}
        \alpha \\ \beta \\ \gamma \\ \delta
    \end{bmatrix}
    =
    \frac{1}{T^7}
    \begin{bmatrix}
        -33600 &  16800T  & -3360T^2 & 280T^3\\
        16800T & -8160T^2 &  1560T^3 & -120T^4\\
        -3360T^2 & 1560T^3 & -280T^4 & 20T^5\\
        280T^3 & -120T^4 &  20T^5 & -\frac{4}{3}T^6
    \end{bmatrix} 
    \begin{bmatrix}
        \Delta p \\ \Delta v \\ \Delta a \\ \Delta j
    \end{bmatrix}
\end{equation}
where the constants $\Delta p$, $\Delta v$, $\Delta a$, $\Delta j$ are defined as,
\begin{equation}\label{eq_apn_delta_pvaj}
    \begin{bmatrix}
        \Delta p \\ \Delta v \\ \Delta a \\ \Delta j
    \end{bmatrix} =
    \begin{bmatrix}
        p_f - p_0 - v_0T -\frac{1}{2}a_0T^2 -\frac{1}{6}j_0T^3\\
        v_f - v_0 - a_0T - \frac{1}{2}j_0T^2\\
        a_f - a_0 - j_0T\\
        j_f - j_0
    \end{bmatrix} 
\end{equation}
Therefore, given the current state $\mathcal{X}_0$ and the desired state $\mathcal{X}_f$, the optimal motion primitive and control input for the fully defined end state can be solved by \eqref{apd_eq_min_s}\eqref{apd_eq_prim_p}\eqref{eq_apn_abcd_1111}\eqref{eq_apn_delta_pvaj}.

\end{document}